\documentclass[10pt]{iopart}

\expandafter\let\csname equation*\endcsname=\relax
\expandafter\let\csname endequation*\endcsname=\relax
\usepackage{amsmath}
\usepackage{iopams}  
\usepackage{float}
\usepackage{txfonts}
\usepackage{array}
\newcolumntype{C}[1]{>{\centering\arraybackslash}p{#1}}

\usepackage{graphicx}
\usepackage{subcaption}  %subcaption
\usepackage{pdfpages}
\usepackage{caption}
\usepackage{cite}
\usepackage{xcolor}
\usepackage{hyperref}
\usepackage{booktabs}
\usepackage{longtable}
\usepackage{array}

\DeclareCaptionFormat{myformat}{\textbf{#1}#2#3}
\begin{document}
\frenchspacing

%\title[Advanced Recognition and Control of Plasma Shape Parameters in HL-3 Using a Novelly Adapted Swin Transformer]{Advanced Recognition and Control of Plasma Shape Parameters in HL-3 Using a Novelly Adapted Transformer}
\title[Physics-Gated Visual Prediction of MARFE on the HL-3 Tokamak]{Physics-Gated Visual Prediction of MARFE on the HL-3 Tokamak}

%\author{Content \& Services Team}

%\ead{customerservices@ioppublishing.org}
%\vspace{10pt}
%\begin{indented}
%\item[]August 2017 (minor update March 2024)
%\end{indented}

\author{Qianyun Dong$^{1,2}$, Rongpeng Li$^{1,*}$, Zongyu Yang$^{2}$, Niannian Wu$^{1,2}$, Fan Xia$^{2}$, Liang Liu$^{2}$, Zhifeng Zhao$^{3}$, Wulyu Zhong$^{2,*}$}
\address{1. Zhejiang University, Hangzhou 310058, China}
\address{2. Southwestern Institute of Physics, Chengdu 610043, China}
\address{3. Zhejiang Lab, Hangzhou 311500, China}
\ead{lirongpeng@zju.edu.cn, zhongwl@swip.ac.cn}

\begin{abstract}
The Multifaceted Asymmetric Radiation From the Edge (MARFE) is a critical plasma instability that often precedes density-limit disruptions in tokamaks, posing a significant risk to machine integrity and operational efficiency. We develop a physics-gated, continuous MARFE monitor for the HL-3 tokamak that outputs a per-frame intensity probability every $2$\,ms, which can potentially be used by the shape-target controller in the plasma control system. Our framework integrates two core innovations: (1) a physics-scored, weighted Expectation-Maximization (EM) pipeline that refines noisy visual labels using $(n_e, T_e, f_G, t)$ as a Bayesian prior, and (2) a continuous-time, physics-gated Neural Ordinary Differential Equation (Neural ODE) backbone whose dynamics are modulated by a sigmoid gate on $f_G$ and $T_e$. Meanwhile, the Neural ODE adopts a $40$\,ms forward forecasting horizon to accommodate the actuator-response budget. On a frozen $140$-shot held-out test set, the proposed method yields a median label-aligned lead time of $+36$\,ms, close to this design horizon. Against a Bi-LSTM baseline trained under the matched protocol, the proposed Neural ODE attains Area Under the Curve (AUC) $=0.981$ and sample-level $F_1=0.840$, compared with AUC $=0.960$ and sample-level $F_1=0.779$ for the baseline. The deployed inference service runs within a $1$-ms control-cycle budget, while new diagnostic samples are generated at the $2$-ms frame cadence.

\vspace{0.3cm} 

\end{abstract}

%\maketitle
\ioptwocol

\section{Introduction}

The central goal of magnetic confinement fusion is to achieve steady, high-performance operation in tokamaks; plasma disruption is the primary obstacle \cite{hender2007mhd}. Among disruptions, density-limit terminations form a dominant family, and they are typically seeded by a \emph{multifaceted asymmetric radiation from the edge} (MARFE) \cite{lipschultz1987review,Lipschultz1984NF,Greenwald1988NF}: a radiative thermal instability that forms on the high-field-side edge when the line-averaged density approaches the Greenwald limit $n_g = I_p / (\pi a^2)$ \cite{Lipschultz1984NF,Greenwald1988NF,Kelly2001PoP}. Reactor-class scenarios such as ITER require sustained operation close to this limit \cite{aymar2002iter,bigot2019iter}. The operation near the Greenwald limit, maintained by either gas puffing or pellet injection, is conducive to density-limit MARFE, so a reliable early MARFE monitor is 
needed independently of the fueling pathway. Active mitigation by adjusting edge conditions and geometry exists \cite{Samm1999JNM,Hollmann2015PoP}, but it cannot be triggered without a reliable early indicator. % The normalised density $f_G = n_e / n_g$ and the core electron temperature $T_e$ are the two principal physics indicators we use throughout this paper.

Existing MARFE-detection approaches fall into three categories. \emph{(i) Bolometer-based tomography} reconstructs the 2-D radiated-power distribution from a small number of line integrals \cite{huber2007improved,Bernert2014RSI,Peluso2019RSI}. However, the inversion is ill-posed and tends to smooth out the localised emissivity features of an early-stage MARFE, limiting its usefulness for early warning. \emph{(ii) Threshold- and morphology-based visible-light imaging} extracts MARFE-like features (bright high-field-side [HFS] region, Hu invariant moments, motion of the radiation centroid) from raw camera streams \cite{Murari2010TPS,Craciunescu2012FST,Albuquerque2011TPS,murari2024control}. But the labels are fragile because divertor glow, wall reflections, gas-puff plumes and post-disruption afterglow easily trip the area threshold \cite{Losada2020NME,Lotte2010RSI,Carr2019RSI,Marini2023RSI}. \emph{(iii) Machine-learning-based visible-light detection} progressively replaces hand-tuned rules. For example, tree-based classifiers track MARFE motion under density-limit conditions on EAST \cite{Hu2023CPB}, while a recent JET image-processing pipeline with a machine-learning back-end detects MARFE earlier than the mode-locked amplitude signal at a reported accuracy of $96.9\%$ \cite{GonzalezGanzabal2024FED}, complementing the earlier OpenCV centroid-tracking approach of \cite{Spolladore2023FED}. Nevertheless, the deep-learning methodologies for disruption prediction  \cite{KatesHarbeck2019Nature,GonzalezGanzabal2024FED,stuart2021petra} require extensive labeled data for training, and there is still no cost-effective way to obtain physics-consistent MARFE labels.

Given the complexity of diagnosing MARFE, the reliability of a visual MARFE monitor is hindered by two reasons. First, limitations in the available diagnostics impose practical constraints on data quality and label reliability. In practice, due to the susceptibility of scene-dependent artifacts, such as bright divertor spots, metallic wall reflections, and gas-puff plumes, simple threshold and morphology heuristics can be easily misled. Therefore, robust labels require physically consistent filtering of noisy visual evidence, and it becomes crucial to build an effective means to stabilize vision-derived labels. In this regard, maximum-likelihood and expectation--maximization (EM) methods \cite{dempster1977maximum,mclachlan2008algorithm} have been applied to noisy fusion inversions, such as bolometric tomography, where statistical priors stabilize inference from imperfect data \cite{Peluso2019RSI,Craciunescu2023PhysScr}.
Second, MARFE dynamics couple multiple evolving parameters (e.g., density, temperature, safety factor, shaping, fueling, and heating) in a non-linear way. Purely data-driven models can fit trends but may extrapolate poorly or violate physics when data are scarce. Incorporating domain knowledge can improve generalization and keep predictions physically plausible \cite{Raissi2019JCP,Willard2022CSUR}. Moreover, because the target is a short-horizon forecast of a fast edge phenomenon, a continuous-time modeling view is natural. Neural ordinary differential equation (ODE) and related controlled ODE frameworks \cite{Chen2018NeurIPS,Rubanova2019NeurIPS,Kidger2020NeurIPS} provide a compact way to evolve latent states under time-varying drives while ensuring consistency with established physical principles, in contrast to discrete-time recurrent networks whose fixed step size is dictated by the sampling rate rather than by the physics itself.

This work aims to develop a physics-gated, continuous MARFE monitor for the HL-3 tokamak, by combining a physics-prior label-refinement pipeline with a continuous-time monitor model designed for closed-loop control feedback. The 
three-stage label refinement pipeline builds high-quality MARFE training targets from noisy camera streams. This process uses a physics-scored and weighted EM algorithm to align visual cues with the underlying plasma state, defined by parameters such as core electron density ($n_{e}$), the normalized density $(f_G = n_e/n_g)$, and core electron temperature ($T_{e}$).
%uses a weighted EM procedure \cite{dempster1977maximum,mclachlan2008algorithm}, in which a per-sample physics score derived from the normalised density $f_G = n_e / n_g$, the core electron temperature $T_e$ and the core electron density $n_e$ and the discharge phase $t$ encodes Greenwald fraction and electron-cooling criteria as a Bayesian prior on the label, and two hard physics gates remove early-time and sub-Greenwald samples deterministically; statistical priors of this kind have previously stabilised noisy fusion inversions \cite{Craciunescu2023PhysScr}. 
Second, to capture the complex dynamics leading to instability, we model a short-horizon Neural ODE monitor \cite{Chen2018NeurIPS,Rubanova2019NeurIPS,Kidger2020NeurIPS}. 
%The prediction model is a physics-gated Neural ODE \cite{Chen2018NeurIPS,Rubanova2019NeurIPS,Kidger2020NeurIPS}: 
Particularly, the Neural ODE evolves a latent state over the $40$-ms prediction horizon, and a sigmoid gate $g(f_G, T_e)$ restricts the dynamics to physically plausible regions, broadly in the spirit of physics-informed learning \cite{Raissi2019JCP,Willard2022CSUR} but without imposing a partial differential equation loss. The label-refinement pipeline is further validated against an expert-reviewed $100$-shot subset by an HL-3 operations physicist. Lastly, we deploy the framework on the HL-3 tokamak and validate its real-time inference capability.
In summary, alongside predicting near-term MARFE worsening with physically consistent dynamics, the proposed physics-informed visual monitor can support the generation of
low-intrusion, geometry-based control targets that are compatible with high-performance, long-pulse operation on devices such as ITER.

\section{Method}

\subsection{Dataset}

All experimental data used in this study are sourced from the HL-3 tokamak, operated by the Southwestern Institute of Physics (SWIP) in China. HL-3 is a medium-sized tokamak with an aspect ratio of $2.8$: plasma current $I_{p}$ = $2.5$–$3$ MA, toroidal field $B$ = $2.2$–$3$ T, major radius $R$ = $1.78$ m, minor radius $a$ = $0.65$ m, and elongation $\kappa\leq1.8$; triangularity $\delta\leq0.5$\cite{duan2022progress}. HL-3 is designed to have a flexible configuration in order to explore multiple divertor configurations. Three heating and current drive (HCD) systems are able to provide a total power of $27$ MW, including $15$ MW of NBI, $8$ MW of electron cyclotron resonance heating (ECRH), and $4$ MW of lower hybrid current drive (LHCD). 

For clarity, a comprehensive list of all symbols and their corresponding descriptions 
is provided in Table~\ref{tab:notations_appendix} in \ref{app:notation}. For each plasma discharge (\emph{shot}), we collect two types of heterogeneous time-series data: visual diagnostics data and zero-dimensional (0-D) plasma parameters. The visual diagnostic data consist of sequential image \emph{frame}s captured by a CCD camera. These image sequences, with a resolution of $640 \times 360$ pixels, serve as the primary source for identifying the spatial location and estimating the morphology and intensity of MARFEs. The 0-D parameters are a set of high-frequency scalar diagnostic signals characterizing the global macroscopic state of the plasma, including environmental and plasma shape parameters. The plasma-state, actuator, and equilibrium-reconstructed parameters cover the real-time core-point electron density and temperature ($n_e$ and $T_e$, both from Thomson scattering), internal inductance ($l_i$), external gas puffing rate (GAS), and injection power from major auxiliary heating systems, including ECRH, LHCD, and neutral beam injection (NBI). Plasma shape parameters include the plasma major radius (R), minor radius (a), vertical position (Z), elongation ($\kappa$), and upper/lower triangularity ($\delta_{u}, \delta_{l}$). These signals are sampled or resampled at $\Delta t=2$\,ms and strictly time-aligned with the visual data, collectively forming the input for our predictive model. %These signals are sampled or resampled at $\Delta t=2$\,ms and strictly time-aligned with the visual data, collectively forming the input for our predictive model. 
To meet real-time constraints, we restrict model inputs to channels whose acquisition and preprocessing latency on our system is less than $1$\,ms, so data delivery does not become the bottleneck. For network inputs, we apply per-channel min--max normalization to $[0,1]$. In particular, $n_e$ and $T_e$ are mapped from the observed ranges $[-3,15]$ and $[-1,13]$ (dataset units) to $[0,1]$. The small negative tail of $n_e$ and $T_e$ comes from baseline-subtraction noise of the upstream Thomson-derived density and temperature signals (typical RMS noise $\sim\!5\%$ for $n_e$ and $\sim\!10\%$ for $T_e$), not from any real negative density or temperature; we keep these mildly negative values in the input range so that the min--max mapping stays continuous across zero. Thresholds used by the physics gate are mapped by the same affine transform so that inputs and thresholds remain in the same space.

Throughout the rest of the paper we use the following operational definition for a \emph{true MARFE} event: a contiguous bright HFS region in the visible-light CCD frame occurring under density-limit conditions, characterized by an elevated Greenwald fraction ($f_G \gtrsim 0.5$) and a simultaneous cooling trend in the available Thomson-derived electron-temperature signal. This combines the canonical visual signature \cite{Lipschultz1984NF,lipschultz1987review} with the two principal 0-D physics indicators ($f_G$, $T_e$) used in the physics prior of Section~\ref{subsec:scoring}. Correspondingly, a MARFE-positive shot is a discharge containing at least one worsening event, and one MARFE-positive shot may contain multiple disjoint events.

The corpus comprises $857$ HL-3 discharges drawn from the disruption-relevant database. Its breakdown by MARFE label and disruption outcome, together with the multi-event statistics relevant to performance, is summarized in Section~\ref{sec:ExperimentalResults}.

\subsection{Physics-Informed MARFE Label Refinement}

\begin{figure}[tbp]
    \centering
    \includegraphics[width=\columnwidth]{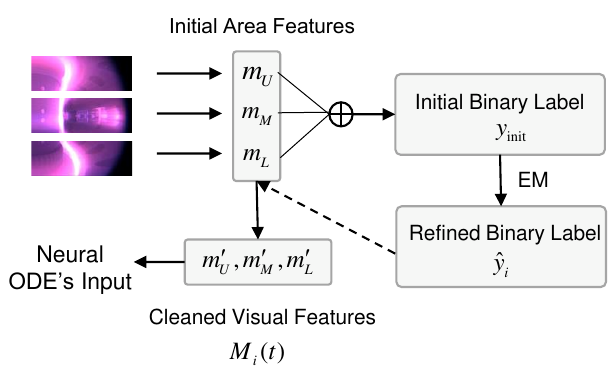} 
    \caption{Data processing pipeline. Initial area features $(m_U, m_M, m_L)$ extracted from raw images yield an initial binary label $y_\mathrm{init}$; the EM algorithm produces a refined label $\hat{y}_i$ used to clean the initial features into the final model inputs $(m'_U, m'_M, m'_L)$.}
    \label{fig:label_features}
\end{figure}

MARFE labels extracted directly from CCD images are often severely contaminated by non-MARFE phenomena such as wall reflections, bright spots from divertor strike points, and gas puffing. To construct a high-quality training dataset, we design and implement a three-stage pipeline to generate robust, physically-consistent pseudo-labels and physically-consistent MARFE data, with the data flow illustrated in Figure ~\ref{fig:label_features}.
First, in the \emph{preliminary feature extraction} stage, raw images are processed to generate \emph{initial area features} (i.e., $m_U$, $m_M$ and $m_L$) and a corresponding \emph{initial binary label} $y_{\text{init}}$. Second, a \emph{physics score} $s_i \in [0, 1]$ is computed for each sample $i$ to quantify the likelihood of MARFE formation based on plasma parameters. Finally, in the \emph{label refinement} stage, we employ a weighted expectation-maximization (EM) algorithm that integrates the physics score $s_i$ as a sample-specific prior to update the \emph{initial binary label} $y_{\text{init}}$. This process yields the final \emph{refined binary label} $\hat{y}_i$, which is used for yielding \emph{cleaned visual features} $(m'_{U}, m'_{M}, m'_{L})$. 

\subsubsection{Preliminary Feature Extraction}\label{sec:primary_feature_extraction}
\begin{figure*}[h!]
    \centering
    \includegraphics[width=0.95\textwidth]{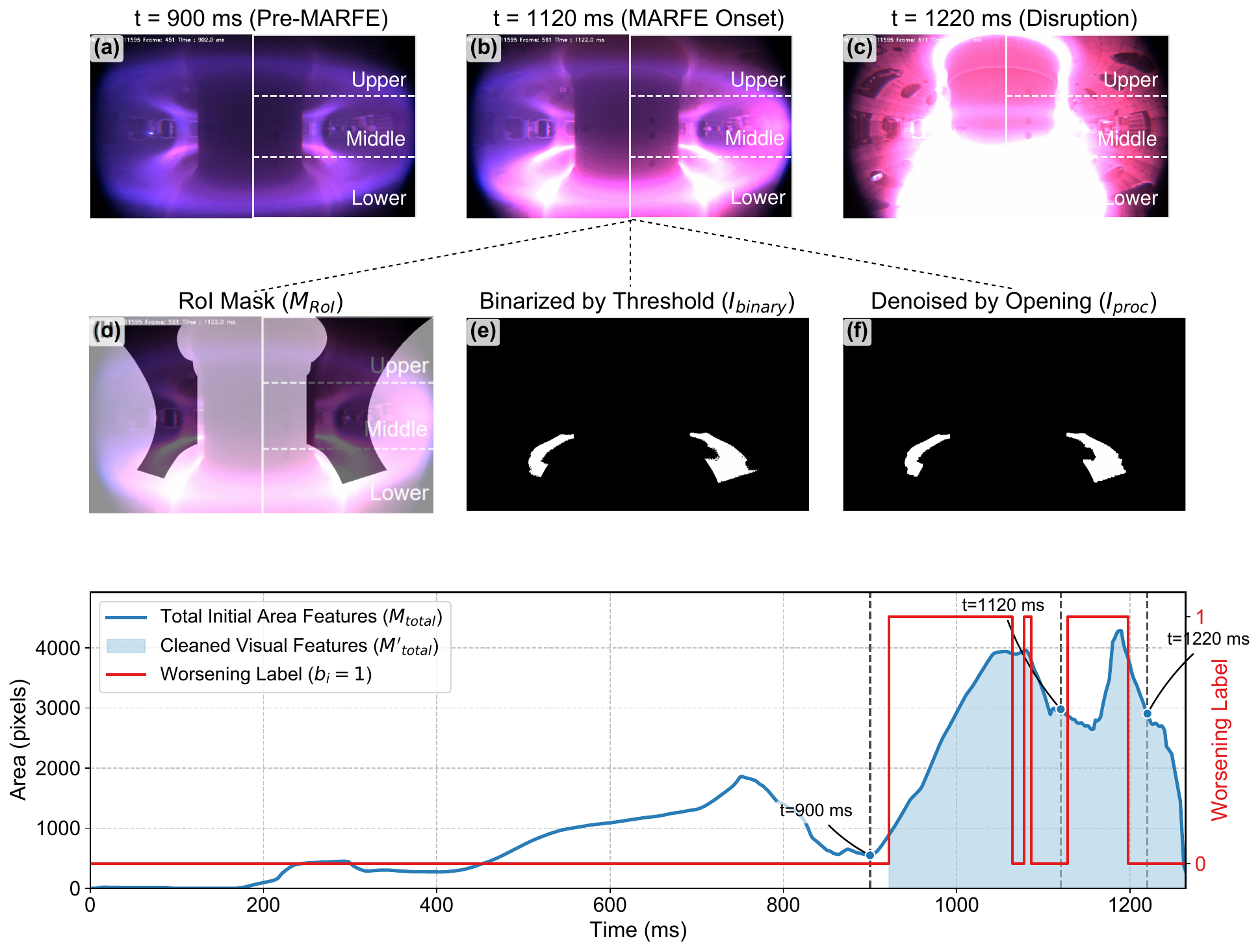}
    \caption{Image processing pipeline for preliminary MARFE feature extraction on shot \#11595. \emph{Top}: CCD frames (a) before MARFE, (b) at onset, (c) at disruption. \emph{Middle}: ROI-masking, binarisation and morphological opening applied to frame (b). \emph{Bottom}: raw vs.\ refined MARFE area signals.}
    \label{fig:marfe_data_processing}
\end{figure*}

% \subsection{Preliminary Feature Extraction}
The first stage of the pipeline leverages a raw CCD image to generate cleaned area features, which correspond to the MARFE intensity within specific zones.
As illustrated in Figure~\ref{fig:marfe_data_processing}, this stage begins by applying a binary region of interest (RoI) mask $M_{\text{RoI}}$ to the grayscale image $I_{\text{gray}}$ and highlighting physically possible regions in the masked image $I_{\text{mask}} = M_{\text{RoI}}\odot I_{\text{gray}}$. High-intensity pixels characteristic of MARFEs are then identified via a brightness threshold $T_{\text{bright}}=220$, yielding a binary image $I_{\text{binary}}$. The value $T_{\text{bright}}=220$ is chosen according to the per-pixel intensity histograms of a labelled subset of MARFE versus background frames, as it lies at the local minimum between the two modes of the bimodal distribution, separating the MARFE-bright population (typically $\ge\! 230$) from the camera background ($\lesssim\! 200$). Afterward, a morphological opening operation with a $5 \times 5$ kernel is applied to remove sensor noise, producing a refined feature map $I_{\text{proc}}$ of MARFE structures.

On this basis, according to the adjacency to the central column in the plasma's HFS,  %which are defined by vertically partitioning the plasma's high-field side (HFS)—the region adjacent to the central column. 
%The zones are designated as upper, middle, and lower.
we partition the processed image $I_{\text{proc}}$ into three distinct poloidal zones: the upper divertor region ($m_U$), the high-field side region ($m_M$), and the lower divertor region ($m_L$). For each zone, the size of the MARFE area (i.e., $m_U$, $m_M$ and $m_L$) is computed according to the summation of non-zero pixels, and correspondingly serve as the initial area features for our prediction model. Using the MARFE area as a zero-order detection metric is consistent with established morphology or imaging-based approaches developed for MARFE identification on JET and other tokamaks~\cite{Murari2010TPS,Craciunescu2012FST,Albuquerque2011TPS}. Compared to \cite{Murari2010TPS,Craciunescu2012FST,Albuquerque2011TPS}, the novel element in our partition is splitting the area into the three poloidal zones $(m_U, m_M, m_L)$ rather than a single aggregated count, which preserves the poloidal-distribution information needed to distinguish a MARFE from generic edge brightening.
% For each zone, the feature is computed as the total MARFE area. These three features serve as direct inputs for the subsequent prediction model. 
We also generate an initial binary label $y_{\text{init}}$ as an initial guess of whether any significant MARFE activity is present in the frame: $y_{\text{init}}=1$ if the aggregated initial area feature $M_{\text{total}} = m_U + m_M + m_L$ exceeds $200$ pixels, and $y_{\text{init}}=0$ otherwise. This binary label is used exclusively within the EM algorithm; if a sample's label is corrected from $1$ to $0$ during refinement, the corresponding feature values ($m_U, m_M, m_L$) for that time step are changed to zero. To maintain clarity, we denote these \emph{cleaned visual features}, which serve as the direct inputs for the subsequent prediction model, as $(m'_{U}, m'_{M}, m'_{L})$.

\subsubsection{Physics Consistency Scoring}\label{sec:physics score}
\label{subsec:scoring}
\begin{figure}[!tb]
    \centering
    \includegraphics[width=0.9\columnwidth]{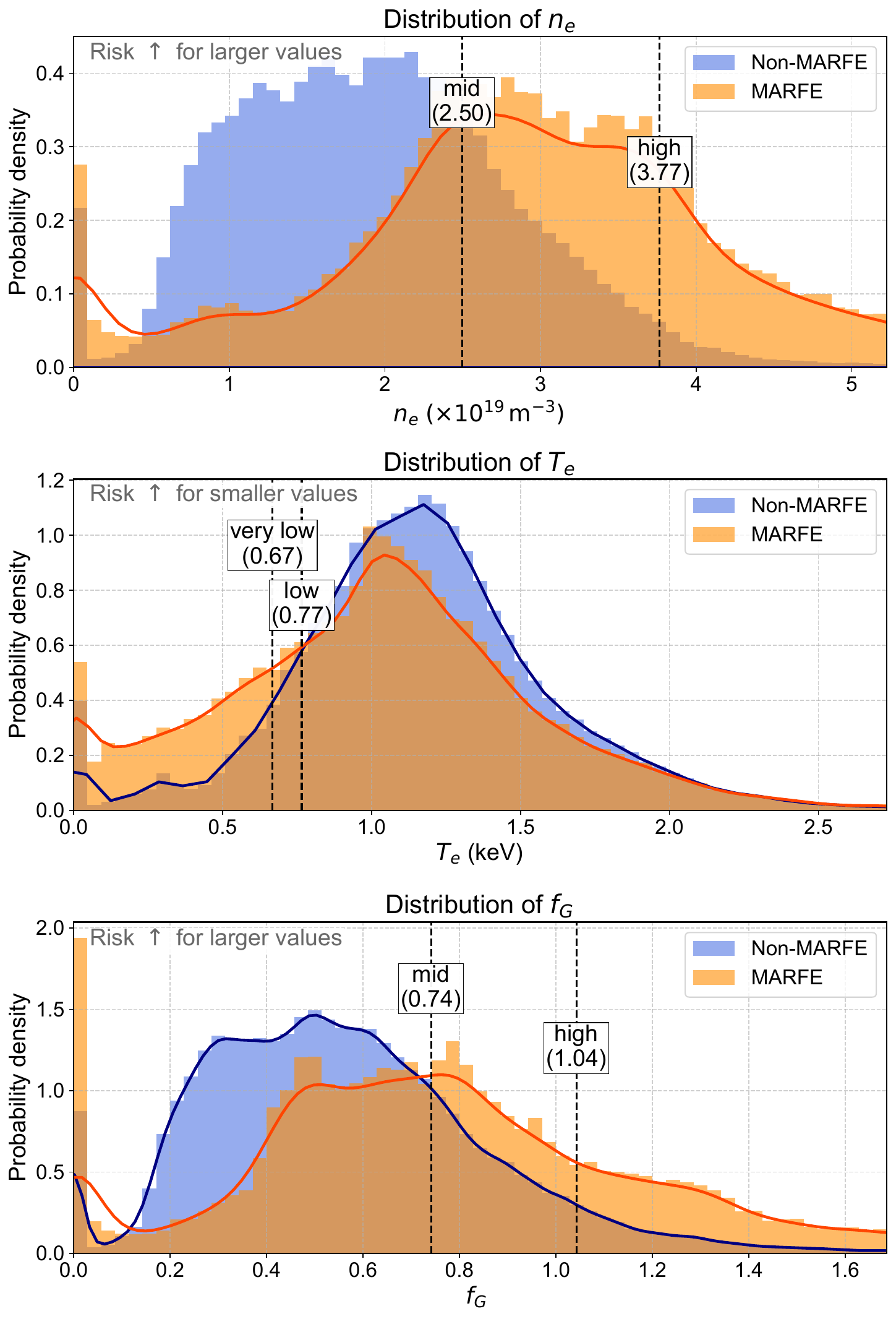} 
    \caption{Probability density of $n_e$, $T_e$ and $f_G$ for MARFE and non-MARFE states. Vertical dashed lines mark the primary and secondary thresholds of Table~\ref{tab:thresholds}.}
    \label{fig:thresholds}
\end{figure}

Since the ``hard" binary label $y_\text{init}$, obtained from the previous stage, contains no measure of physical certainty, its accuracy is limited due to the inherent noise in visual diagnostics. To address this, we incorporate physical prior knowledge that can quantify how conducive the current plasma state is to MARFE formation. For this purpose, we construct a physics consistency scoring function $s_i \in [0, 1]$ for each sample $i$, by transforming several MARFE worsening-related 0-D physical parameters. Here, the normalized density $f_G= n_e / n_g$ is the principal density-related indicator, since $n_g = I_p / (\pi a^2)$ absorbs the shot-to-shot variation of the plasma current $I_p$ and minor radius $a$ that controls the Greenwald limit, while $n_e$ enters only as a secondary, absolute-magnitude signal in the physics score. Besides, the core electron temperature ($T_e$) is also involved. 
These underlying parameters, $I_p$ and $a$, are available in our dataset. We note that the conventional Greenwald fraction is defined using the line-averaged density. In this work, however, $f_G$ is computed using the real-time core-point density $n_e$, because the real-time deployment does not simultaneously provide both the line-averaged density and the core-point density required by the present monitor. While this approach may result in $f_G$ values that are comparatively higher than those reported in literature using line-averaged density~\cite{Lipschultz1984NF,lipschultz1987review}, the parameter's trend and its strong correlation with MARFE onset remain robust and effective for prediction. %To ensure these evaluations are objective and optimally tuned to our dataset, the critical thresholds are determined automatically using a data-driven approach.

For each parameter, we employ the Youden's J statistic from a receiver operating characteristic (ROC) curve analysis on the initial noisy labels ($y_{\text{init}}$) to find the optimal threshold that separates the MARFE and non-MARFE distributions~\cite{youden1950index}. Figure~\ref{fig:thresholds} visualizes the corresponding distributions. On this basis, thresholds defining ``high'' or ``low'' levels are derived from the quantiles (e.g.\ 75th percentile for parameters like $n_e$ where higher values indicate greater risk) of the MARFE-positive distribution. While these thresholds, statistically annotated in Figure~\ref{fig:thresholds} and listed in Table~\ref{tab:thresholds}, are derived from the initially noisy visual labels, this data-driven approach serves as a robust bootstrapping mechanism. The thresholds in Table~\ref{tab:thresholds} are derived for HL-3 from a labelled subset of its discharges; transferring the method to another tokamak requires re-deriving the primary thresholds via Youden's $J$ on a small labelled bootstrap ($50$--$100$ shots) of the new device.
On top of these thresholds, the score $s_i$ is built in two stages.

The first stage applies two physics hard gates (i.e., time $t$ and $f_G$) to down-weight samples for which a MARFE is highly unlikely under the present HL-3 operating scenarios. The first sets $s = 0.01$ for $t < t_\mathrm{cut} =  300$\,ms, before the plasma edge becomes confined; the prior is kept strictly positive for log-prior numerical stability. The second sets $s = 0.01$ for sub-Greenwald plasmas at $t \ge  t_\mathrm{cut}$ with $f_G < 0.40$, where the density is too low to drive the radiative thermal instability. This $0.40$ cutoff matches the 5th percentile of $f_G$ at MARFE onset across our $216$-event manually-annotated subset, and lies below the canonical MARFE-onset range $f_G \approx 0.40$--$0.55$ reported across earlier tokamaks~\cite{Lipschultz1984NF,lipschultz1987review}. Here, $f_G$ is a device-normalized density indicator, whereas $t$ is used only as a shot-phase proxy. On HL-3, $t$ carries useful operation-schedule priors: across the full $857$-shot corpus, the MARFE class has median $t=2002$\,ms, whereas the non-MARFE class has median $t=1102$\,ms. This phase information helps distinguish phase-dependent visual artifacts, such as post-disruption afterglow, from physically plausible MARFE states; however, the use of $t$ should be regarded as a device-specific optimization rather than a universally transferable physical input.

For every sample that passes both gates, the second stage computes $s_i$ as a weighted sum of indicator functions:
\begin{equation}
\begin{split}
s_i = & \min\Big\{1.0,  \; 0.2 \cdot \mathbb{I}(n_{e,i} > n_e^{\text{mid}}) + 0.1 \cdot \mathbb{I}(n_{e,i} > n_e^{\text{high}}) \\
& + 0.2 \cdot \mathbb{I}(T_{e,i} < T_e^{\text{mid}}) + 0.1 \cdot \mathbb{I}(T_{e,i} < T_e^{\text{low}}) \\
& + 0.3 \cdot \mathbb{I}(f_{G,i} >  f_G^{\text{mid}}) + 0.1 \cdot \mathbb{I}(f_{G,i} > f_G^{\text{high}}) \Big\},
\end{split}
\label{eq:scoring_function}
\end{equation}
\noindent where $\mathbb{I}(\cdot)$ is the indicator function. The weights (e.g., $0.3$ for $f_G > f_G^{\text{mid}}$) are chosen based on the known physical importance of each parameter in MARFE formation~\cite{lipschultz1987review, stroth2022model}; the $\min(\cdot, 1)$ is a defensive cap that keeps $s_i \in [0,1]$, rarely active in practice since the current weights sum to $1.0$. This formulation yields a score $s_i$ approaching $1$ for plasma states highly prone to MARFE and approaching $0$ for MARFE-unlikely states. 
The identification of highly susceptible regions forms the basis for calculating the physics score $s_i$ and allows the subsequent EM algorithm to refine labels with a physically-grounded prior, even in the presence of initial label noise. 

\begin{table}[tb]
    \centering
    \caption{Empirically determined thresholds for physics-based prior calculation. The ``$\text{mid}$'' superscript denotes the primary threshold (Youden's $J$ cut), and ``$\text{high}$''/``$\text{low}$'' denote the secondary thresholds (distribution percentile). The ``Source'' column indicates whether the threshold is the Youden's $J$ optimal cut on the ROC of the initial noisy labels (primary thresholds) or the $25$th / $75$th percentile of the MARFE-positive distribution (secondary thresholds).}
    \begin{tabular}{lrl}
        \br
        \textbf{Symbol} & \textbf{Value} & \textbf{Source} \\
        \mr
        $n_e^{\text{mid}}$  & $2.496\times10^{19}\,\text{m}^{-3}$ & Youden's $J$ \\
        $n_e^{\text{high}}$ & $3.765\times10^{19}\,\text{m}^{-3}$ & $P_{75}$ \\
        \addlinespace
        $T_e^{\text{mid}}$  & $0.766\,\text{keV}$ & Youden's $J$\\
        $T_e^{\text{low}}$  & $0.668\,\text{keV}$ & $P_{25}$ \\
        \addlinespace
        $f_G^{\text{mid}}$  & $0.741$ & Youden's $J$ \\
        $f_G^{\text{high}}$ & $1.043$ & $P_{75}$ \\
        \br
    \end{tabular}
    \label{tab:thresholds}
\end{table}

\subsubsection{Weighted EM-based Label Refinement}\label{sec:em_refinement}
\begin{figure}[tbp]
    \centering
    \includegraphics[width=\columnwidth]{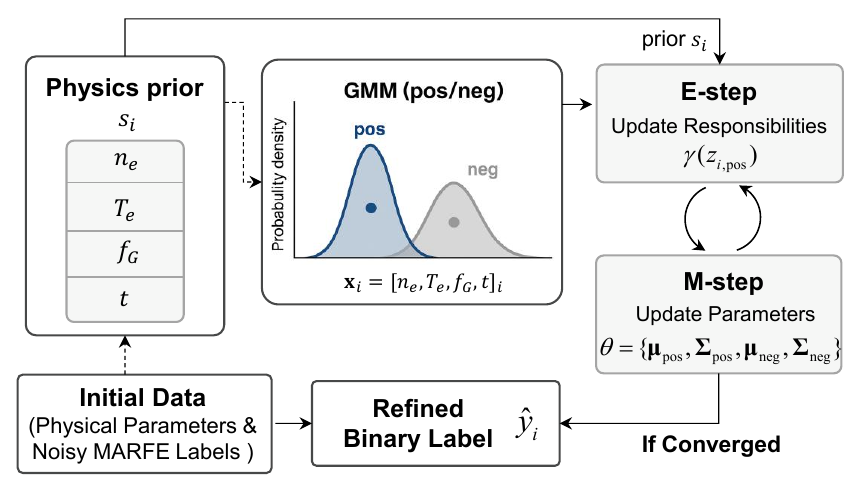} 
    \caption{Schematic of the physics-informed weighted EM label refinement. The physics prior $s_i$ enters the E-step as a sample-specific empirical physics prior on the latent MARFE state; $y_\mathrm{init}$ does not enter the E- or M-step and is used only as the binary reference whose flips define the cleaned set.}
    \label{fig:em_refinement_workflow}
\end{figure}

The core of our model is to refine the noisy visual labels by leveraging the physics-based prior probability $s_i$. The EM/GMM parameters are fitted offline once on the historical HL-3 corpus and then frozen before any supervised predictor training or model comparison. In real-time operation, the fitted parameters are kept fixed and only the frame-wise posterior inference step is evaluated online.
We frame this as a parameter estimation problem for a latent variable, where the true MARFE state is unobservable. Naturally, we resort to the canonical \emph{EM algorithm}~\cite{dempster1977maximum, mclachlan2008algorithm}, by modeling the physics feature vector $\mathbf{x}_i = [n_e, T_e, f_G, t]_i$ as samples from a Gaussian mixture model (GMM), where the true MARFE ($z_i=\text{pos}$) and non-MARFE ($z_i=\text{neg}$) states correspond to two distinct Gaussian components. For simplicity, we assume conditional independence between features, resulting in diagonal covariance matrices. Although $f_G=n_e/n_G$ is correlated with $n_e$ by construction, we retain both variables because $f_G$ carries shot-normalized Greenwald-limit information through $I_p$ and $a$, whereas $n_e$ provides the absolute density scale. The diagonal-covariance assumption is therefore used as a tractable approximation rather than as a claim of strict statistical independence. As schematically illustrated in Figure~\ref{fig:em_refinement_workflow}, the EM algorithm iteratively optimizes the GMM parameters $\theta = \{\boldsymbol{\mu}_{\text{pos}}, \boldsymbol{\Sigma}_{\text{pos}}, \boldsymbol{\mu}_{\text{neg}}, \boldsymbol{\Sigma}_{\text{neg}}\}$. Here, $\boldsymbol{\mu}$ and $\boldsymbol{\Sigma} = \text{diag}(\boldsymbol{\sigma}^2)$ denote the mean vectors and the diagonal covariance matrices, respectively, where $\boldsymbol{\sigma}^2$ is the variance vector. 
%\paragraph{Initialization Strategy}
Since a robust initialization of the GMM parameters is crucial for convergence, we leverage the physics prior $s_i$ to guide this process. To enhance the statistical robustness, we take the top $10\%$ of the $s_i$ distribution as the high-prior seed set (a confident MARFE proxy) and the bottom $10\%$ as the low-prior seed set (a confident non-MARFE proxy). If at least two samples exist in the high-prior seed set, their feature mean and standard deviation are used to initialize the positive component's parameters, $\boldsymbol{\mu}_{\text{pos}}^{(0)}$ and $\boldsymbol{\sigma}_{\text{pos}}^{(0)}$. A similar procedure is applied to initialize the negative component using the low-prior seed set. %If either set contains fewer than two samples, we initialize the corresponding component's parameters using a small, random subset of the data. 
This strategy ensures that the model is guided by the most physically plausible data points. Afterward, we alternately iterate the EM algorithm between the expectation-step (E-step) and the maximization (M-step) until convergence.

For the $k$-th \emph{E-Step}, 
%In iteration $t$, 
based on the current parameter estimates $\theta^{(k)}$, we calculate the posterior probability, or responsibility $\gamma(z_{i,\text{pos}})$, that data point $\mathbf{x}_i$ belongs to the true MARFE. Because $s_i$ is computed from the same diagnostic vector $\mathbf{x}_i$, we interpret it as a sample-specific empirical physics prior rather than an independent prior. This weight modulates the GMM likelihoods as the likelihood of $\mathbf{x}_i$ under each component is given by the multivariate Gaussian probability density function (PDF):
\begin{subequations}
\begin{align}
    L_{i,\text{pos}}^{(k)} &= \mathcal{N}(\mathbf{x}_i | \boldsymbol{\mu}_{\text{pos}}^{(k)}, \boldsymbol{\Sigma}_{\text{pos}}^{(k)}); \\
    L_{i,\text{neg}}^{(k)} &= \mathcal{N}(\mathbf{x}_i | \boldsymbol{\mu}_{\text{neg}}^{(k)}, \boldsymbol{\Sigma}_{\text{neg}}^{(k)}).
\end{align}
\end{subequations}
Meanwhile, the responsibility $\gamma(z_{i,\text{pos}})$ is computed as:
\begin{equation}
    \gamma(z_{i,\text{pos}}) = \frac{s_i L_{i,\text{pos}}^{(k)}}{s_i L_{i,\text{pos}}^{(k)} + (1 - s_i) L_{i,\text{neg}}^{(k)}}
\end{equation}
To avoid numerical underflow, all calculations involving likelihoods are performed in log-space using the log-sum-exp trick~\cite{blanchard2021accurately}.

For \emph{M-Step}, contingent on the responsibility $\gamma(z_{i,\text{pos}})$, we perform a tempered update, so as to prevent oscillations and improve convergence. First, we compute the target parameters as weighted sample statistics:
\begin{subequations}
\begin{align}    \boldsymbol{\mu}_{\text{target},\text{pos}}^{(k+1)} &= \frac{\sum_{i=1}^{N} \gamma(z_{i,\text{pos}}) \mathbf{x}_i}{\sum_{i=1}^{N} \gamma(z_{i,\text{pos}})}; \\
    (\boldsymbol{\sigma}_{\text{target,pos}}^{2})^{(k+1)} &= \frac{\sum_{i=1}^{N} \gamma(z_{i,\text{pos}}) (\mathbf{x}_i - \boldsymbol{\mu}_{\text{target},\text{pos}}^{(k+1)})^2}{\sum_{i=1}^{N} \gamma(z_{i,\text{pos}})},
\end{align}
\end{subequations}
where the square in the variance calculation is element-wise. Then, we update the current parameters for the positive component as:
\begin{subequations}
\begin{align}
    \boldsymbol{\mu}_{\text{pos}}^{(k+1)} & = (1 - \alpha) \boldsymbol{\mu}_{\text{pos}}^{(k)} + \alpha \boldsymbol{\mu}_{\text{target},\text{pos}}^{(k+1)}; \\
    (\boldsymbol{\sigma}_{\text{pos}}^{2})^{(k+1)} & = (1 - \alpha) (\boldsymbol{\sigma}_{\text{pos}}^{2})^{(k)} + \alpha (\boldsymbol{\sigma}_{\text{target,pos}}^{2})^{(k+1)},
\end{align}
\end{subequations}
where the learning rate $\alpha=0.5$. 
Analogously, the parameters for the negative component can be updated. To prevent numerical instability, the standard deviation for each feature is clamped to a minimum value (e.g., $\boldsymbol{\sigma}^{(k+1)} \leftarrow \max(\boldsymbol{\sigma}^{(k+1)}, 10^{-3})$) after the update. 

%标签判定
Once the EM algorithm reaches convergence, the final posterior probability $\gamma(z_{i,\text{pos}})$ provides a robust, continuous-valued label representing the refined probability of a true MARFE event. Specifically, the refined binary label $\hat{y}_i$ can be obtained as:
\begin{equation}
\label{eq:label_definition}
    \hat{y}_i =
    \begin{cases}
        1 \text{ (MARFE)}, & \text{if } \gamma(z_{i,\text{pos}}) > \gamma_{\text{thre}}; \\
        0 \text{ (non-MARFE)}, & \text{otherwise}.
    \end{cases}
\end{equation}
In practice we set $\gamma_{\text{thre}} = 0.5$ to balance recall and precision. Notably, because the GMM feature vector $\mathbf{x}_i = [n_e, T_e, f_G, t]_i$ includes the discharge phase $t$, transferring the method to a device with a substantially different operational window, such as the longer pulses of EAST or ITER, requires re-fitting the GMM on that device's discharge statistics.

\begin{figure}[!tb]
    \centering
    \includegraphics[width=\columnwidth]{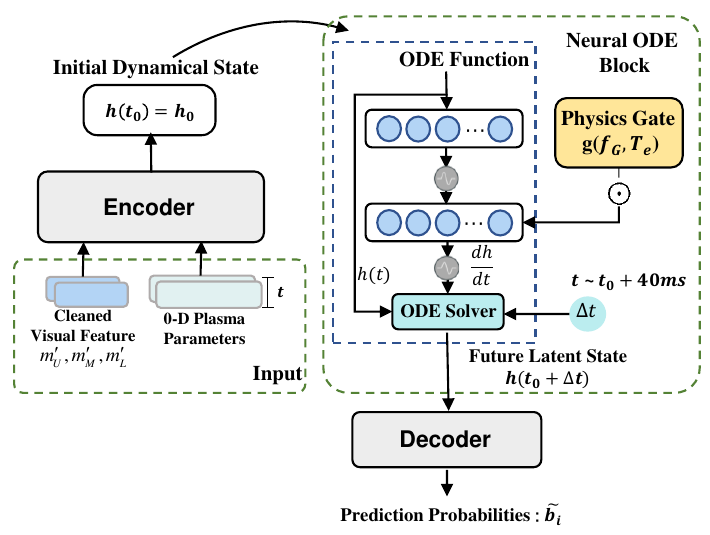}
    \caption{
    The overall architecture of the physics-gated MARFE prediction framework. The \textbf{Neural ODE Model}, which encodes historical time series data, evolves the system's latent state continuously with a physics-gated Neural ODE, and decodes the future state to predict MARFE worsening.
    }
    \label{fig:model_framework}
\end{figure}

As mentioned earlier in Section \ref{sec:primary_feature_extraction}, based on $\hat{y}_i$, we can automatically clean the visual features. For any time index $i$, the regional areas ($m_U, m_M, m_L$) will be updated as:
\begin{equation}
    (m'_U, m'_M, m'_L)_i =
    \begin{cases}
        (m_U, m_M, m_L)_i, & \text{if } \hat{y}_i = 1; \\
        (0, 0, 0),         & \text{if } \hat{y}_i = 0.
    \end{cases}
\end{equation}
In other words, this produces a label–feature set that is physically consistent with reduced error propagation. % Using this cleaned dataset, our goal is not only to detect existing MARFEs but also to forecast short-horizon worsening, which provides a usable time window for mitigation.

\subsection{Physics-Gated Neural ODE-based Prediction Model}

%  We therefore define a forward-looking target, the ``MARFE Worsening Label,'' from the cleaned time series, and then specify the input organization and the architecture of the proposed physics-gated prediction model.

\subsubsection{Model Input}
To accurately capture the state of the plasma and the conditions leading to MARFE formation, our model utilizes a wide array of diagnostic signals. These signals are treated as a multivariate time series. The inputs, detailed in Table~\ref{tab:model_inputs}, include both global plasma parameters (e.g., plasma current $I_p$, electron density $n_e$ and plasma shape parameters) and the aforementioned cleaned visual features ($m'_U, m'_M, m'_L$), eventually being concatenated as a $D$-dimensional vector $\mathbf{a} \in \mathbb{R}^D$. %This combination of these features provides a holistic view of the plasma dynamics, which is crucial for robust prediction.
For the model to effectively learn temporal dependencies, we structure the input data using a sliding window approach. With a data sampling rate of $\Delta t=2$ ms, we use a time window of $40$ ms. Therefore, at any given time $t$\footnote{Note $t$ is introduced for notational convenience and is not fed into the model.}, the model receives a sequence of the past $T=20$ time steps, denoted as $\mathbf{A}(t) = \{\mathbf{a}(t-T+1), \dots, \mathbf{a}(t)\}$, where $\mathbf{a}(\tau) \in \mathbb{R}^D$ is the vector of all $D$ input features at time $\tau$. All input features are normalized using min-max normalization to ensure they are on a comparable scale, which is essential for stable training of neural networks. The $40$\,ms window length itself is set by the lead time the HL-3 plasma control system needs for activating mitigation actuators (gas-puff modulation, ECRH input, current-ramp adjustment) before MARFE worsening triggers a disruption. These actuators have effective response time (i.e., tens of milliseconds), and $40$\,ms also remains short relative to typical MARFE-to-disruption development time reported in classical work~\cite{Lipschultz1984NF,lipschultz1987review}, so the prediction stays preemptive rather than coincident with the disruption.

\subsubsection{Prediction Target: MARFE Worsening Label}
\label{subsec:label_definition}

% Because the worsening label below is derived from the EM-cleaned visual trajectory, its reliability depends on the upstream EM-refined state label $\hat{y}_i$. We therefore do not treat $\hat{y}_i$ as an infallible ground truth. Instead, it is used as an expert-audited operational reference: the independent $100$-shot audit in Section~\ref{sec:label_pipeline_validation} shows that the EM refinement substantially improves the physical consistency of the visual label, increasing precision from $0.45$ to $0.67$ and $F_1$ from $0.62$ to $0.78$ while retaining high recall. On this basis, Eq.~\eqref{eq:worsening_label} defines $b_j(t)$ from the EM-cleaned area signals $(m'_U,m'_M,m'_L)$ rather than from the raw thresholded areas. 
The core task of our model is to provide a timely warning before a MARFE event becomes severe. To achieve this, for each sampled $\mathbf{A}(t)$, we introduce a forward-looking, binary ``MARFE worsening" label, $b_j(t) \in \{0, 1\}$, for each poloidal zone $j$ (i.e., upper, middle, or lower). This label is set to $1$ if the related MARFE condition is about to intensify significantly. Since even after the EM-based label refinement, the refined $\hat{y}_i$ still cannot be considered an infallible ground truth. Therefore, we resort to a sustained, gradual growth $G_j(t)$ of features $m'_j$ (i.e., $m'_U$, $m'_M$, or $m'_L$) in the recent past followed by a notable jump $\Delta m'_j(t) = m'_j(t+40\text{ms}) - m'_j(t)$ in the future $40$ ms (or equivalently $20$ time steps), or an extremely large and abrupt jump $\Delta m'_j(t)$ in the future $40$ ms, regardless of its previous trend~\cite{Lipschultz1984NF, adams2007bocd}. The label $b_j(t)$ is forward-looking and does not depend on the disruption outcome.
% this label is derived from the time series of the cleaned visual features for each region, $M_i(t)$, which corresponds to the features $(m'_U, m'_M, m'_L)$ generated in our data preprocessing pipeline. A MARFE is considered to be ``worsening" at time $t$ if its area is projected to increase significantly over the next $40$ ms (i.e., $20$ time steps). We define "significant increase" based on two physically meaningful scenarios:
% \begin{enumerate}
%     \item A sustained, gradual growth in the recent past followed by a notable jump in the near future.
%     \item An extremely large and abrupt jump in the near future, regardless of its previous trend.
% \end{enumerate}
% To formalize this, we calculate the future change in MARFE area as $\Delta M_i(t) = M_i(t+40\text{ms}) - M_i(t)$ and the recent growth trend, $G_i(t)$, as the slope of a linear regression over the past 40 ms of data. 
Mathematically, the label $b_j(t)$ for training can be written as:
%The ``worsening" label $b_i(t)$ is then assigned as follows:
\begin{equation}
\label{eq:worsening_label}
b_j(t) = 1 \quad \text{if} \quad
\begin{cases}
    (\Delta m_j^{\text{inst}}(t) > \theta_j^{\text{inst}}) \land (G_j(t) > 0)  \\
    \text{or} \\
    (\Delta m_j^{\text{inst}}(t) > c \cdot \theta_j^{\text{inst}}), 
\end{cases}
\end{equation}
The total-region target is defined as the logical union of the three regional targets:
\[
b_{\mathrm{total}}(t)=\max_{j\in\{U,M,L\}} b_j(t).
\]
%The independently learned total-region head $\hat b_{\mathrm{total}}(t)$ is used as the scalar alarm probability $p(t)$ in all subsequent evaluations.
Here, $G_j(t)$ is computed as the slope of a linear regression over the past 40 ms of data and $\theta_j^{\text{inst}}$ is a pre-defined threshold for a ``notable jump" specific to each region $j$, determined empirically from the training data. The constant $c > 1$ (we use $c=1.5$) ensures that only exceptionally large increases trigger the second condition. The value $c=1.5$ requires the ``sudden jump'' threshold to exceed the ``general jump'' threshold by $50\%$, an empirical choice consistent with the typical separation observed between MARFE-onset events and the slower flat-top area drift in our labelled subset. This dual-criteria definition makes our prediction target robust, allowing the model to flag both developing and sudden-onset MARFE events.

\subsubsection{Physics-Gated Neural ODE Model Architecture}

\begin{table}[t]
    \centering
    \caption{Prediction model inputs with abbreviations, brief notes, and units.}
    \label{tab:model_inputs}
    \renewcommand{\arraystretch}{1.1}
    \begin{tabular}{@{}lllc@{}}
    \br
    Input & Brief note & Units  \\
    \mr
    $I_p$ & plasma current & kA  \\
    $a$ & minor radius & m  \\
    $\kappa$ & elongation & --  \\
    $\delta_u$ & upper triangularity & --  \\
    $\delta_l$ & lower triangularity & --  \\
    $R$ & major radius & m  \\
    $Z$ & vertical position & m  \\
    $l_i$ & internal inductance & --  \\
    $P_{\mathrm{NBI}}$ & NBI power & MW  \\
    $P_{\mathrm{ECRH}}$ & ECRH power & MW  \\
    $P_{\mathrm{LHCD}}$ & LHCD power & MW  \\
    $n_e$  & core electron density & $10^{19}\,\mathrm{m}^{-3}$ \\
    $f_G$ & Normalized density ($n_e/n_g$) & --  \\
    $T_e$  & core electron temperature & keV  \\
    %$\tau_{\mathrm{SMBI}}$ & SMBI pulse width & ms  \\
    $m'_U$ & Cleaned MARFE features (upper) & --  \\
    $m'_M$ & Cleaned MARFE features (middle) & --  \\
    $m'_L$ & Cleaned MARFE features (lower) & --  \\
    \br
    \end{tabular}
\end{table}

With the input features and the worsening label $b_i(t)$, we can now specify the model architecture. While a standard approach for such a time-series forecasting task is a Recurrent Neural Network (RNN), its discrete-time nature is ill-suited to modeling the continuous evolution inherent to plasma physics. RNNs force the system's dynamics into fixed, artificial time steps ($\Delta t$), which can fail to capture critical, fast-evolving phenomena that occur between sampling intervals and make the model's performance dependent on the chosen sampling rate.
% 有什么不足，补一句话
To better capture these dynamics, we propose a hybrid Neural ODE architecture~\cite{Chen2018NeurIPS}, which  first uses a \emph{sequence encoder} to learn a robust latent representation $\mathbf{h}_0$ of the plasma's recent history $\mathbf{A}(t)$. Afterward, it evolves this state via the Neural ODE. Concretely, the sequence encoder is a $2$-layer Bi-LSTM with hidden dimension $128$ per direction, while the ODE block consists of two $2$-layer fully-connected networks with Tanh activation, integrated from $t_0=0$ to $t=40$\,ms by a fixed-step RK4 solver~\cite{butcher2016numerical}.

%The process begins with a \textbf{Sequence Encoder}. The input sequence $\mathbf{A}_t$ is processed by the encoder to effectively capture temporal patterns and dependencies from the historical data. This is crucial for understanding the context of plasma evolution. The encoder processes the entire sequence and produces a final hidden state, $\mathbf{h}_0$. This vector $\mathbf{h}_0$ serves as a compressed, fixed-length latent representation of the plasma's recent history up to time $t$.
%Subsequently, this latent vector $\mathbf{h}_0$ is used as the initial condition for the \emph{neural ODE for latent dynamics}. As Plasma evolution is a continuous-time process, this module is designed to reflect this underlying physics. 
Let $\mathbf{u}(t) := (T_e, n_e, f_G, m'_U, m'_M, m'_L)\big|_t$ denote the instantaneous plasma and visual state at time $t$ (the last six rows of Table~\ref{tab:model_inputs}), while $t$ is not fed into the model. The Neural ODE learns the continuous dynamics of the latent state $\mathbf{h}(t)$ (where $\mathbf{h}(t_0) = \mathbf{h}_0$) governed by the differential equation with a \emph{physics-guided gating mechanism}:
\begin{equation}
\label{eq:ode}
    \frac{d\mathbf{h}(t)}{dt} = f_{\theta}(\mathbf{h}(t), \mathbf{u}) + g(f_G, T_e)\, f_{\phi}(\mathbf{h}(t), \mathbf{u})
\end{equation}
Here $f_{\theta}$ is the data-driven base dynamics, which sees both the latent state $\mathbf{h}$ and the current plasma state $\mathbf{u}$ and is responsible for the history-aware part of the evolution. The residual term $f_\phi$ is a physics-gated correction with the same state and input variables, but its contribution is modulated by the MARFE-onset gate $g(f_G,T_e)$, which is implemented as a sigmoid function and activates as plasma conditions approach a critical MARFE-prone state:
\begin{equation}
    g(f_G, T_e) = \sigma \left( k_n (f_G - f_G^{\text{thr}}) - k_T (T_e - T_e^{\text{thr}}) \right)
\end{equation}
where $f_G^{\text{thr}}$ and $T_e^{\text{thr}}$ are the primary thresholds $f_G^{\text{mid}}$ and $T_e^{\text{mid}}$ derived from the statistical analysis of Section~\ref{subsec:scoring}. When conditions are safe ($g \approx 0$), the dynamics reduce to the base term $f_\theta$; as conditions approach the MARFE boundary ($g \to 1$), the gate admits the physics correction $f_\phi(\mathbf{h},\mathbf{u})$, which captures the MARFE-conditioned correction to the latent dynamics. A further discussion on the physics gate mechanism is given in Section~\ref{sec:phy_gate}. By integrating Eq.~(\ref{eq:ode}) from initial time $t_0$ to $t_0 + \Delta t$, the final state $\mathbf{h}(t_0 + \Delta t)$ represents the predicted future plasma state in the latent space. 

This gated decomposition lets the model retain a general history-aware evolution through $f_\theta$ while adding an explicitly MARFE-conditioned correction through $g(f_G,T_e)f_\phi$. Thus, the physics prior enters the dynamics as an architectural modulation rather than as a hard differential-equation constraint.

Finally, the evolved latent state $\mathbf{h}(t_0 + \Delta t)$ is passed to the \emph{classifier}, which consists of a $2$-layer fully-connected network with GELU activation and a sigmoid output. This produces four sigmoid probabilities: three regional heads $\hat{b}_{i,U}$, $\hat{b}_{i,M}$, $\hat{b}_{i,L}$ and one independently learned total-region head $\hat{b}_{i,\mathrm{total}}$. In the results below, the scalar alarm probability is defined as $p(t)=\hat{b}_{\mathrm{total}}(t)$, and the $N$-frame persistence rule is applied to this total-region probability stream.

\subsubsection{Loss Function and Training}
The model is trained end-to-end to minimize a loss function that accounts for the four prediction tasks comprising the upper, middle, lower, and total-region MARFE-worsening targets. We employ an uncertainty-weighted binary cross-entropy loss~\cite{kendall2018multi}, which allows the model to automatically learn to balance the importance of each region. This is beneficial because the signal-to-noise ratio and predictability can vary significantly between regions. The average loss $\mathcal{L}$ over a mini-batch is formulated as:
\begin{equation}
    \mathcal{L} = \sum_{i\in \text{batch}} \sum_{j\in\{\text{upper},\text{middle},\text{lower},\text{total}\}} \left[ \frac{1}{2\sigma_{j}^{2}}\mathcal{L}_{BCE}(b_{i,j}, \hat{b}_{i,j}) + \ln \sigma_{j} \right]
\end{equation}
where $\mathcal{L}_{\text{BCE}}(b_{i,j}, \hat{b}_{i,j}) = -[b_{i,j} \ln(\hat{b}_{i,j}) + (1-b_{i,j})\ln(1-\hat{b}_{i,j})]$ is the binary cross-entropy for task $j$, and $\sigma_j$ is a learnable parameter representing the homoscedastic uncertainty for that task's prediction. The $1/(2\sigma_j^2)$ term adaptively weights the loss for each zone, while the $\ln \sigma_j$ term acts as a regularizer to prevent the uncertainties from growing infinitely~\cite{kendall2018multi}.

The entire model, including the parameters of the sequence encoder, the neural networks $f_{\theta}$ and $f_{\phi}$, the physics gate, and the uncertainty weights $\sigma_j$, is trained using the Adam optimizer, a learning rate of $10^{-3}$ decayed exponentially by $0.95$ per epoch, batch size $512$, and a sliding-window length of $20$ samples ($=40$\,ms). Gradients are computed via backpropagation through the ODE solver, for which we use the efficient adjoint sensitivity method ~\cite{Chen2018NeurIPS, serban2005cvodes}. Training is capped at $30$ epochs with early stopping on the validation $F_1$. We use early stopping on the held-out validation set to prevent overfitting; the test set is reserved for the final evaluation reported in Section~\ref{sec:ExperimentalResults} and is never seen during training or model selection.

\section{Experimental Results and Analysis}
\label{sec:ExperimentalResults}
We evaluate the proposed physics-gated visual prediction method on the HL-3 dataset of $857$ shots with IDs in range \#4400--\#12000. Particularly, we retain only $701$ shots with complete diagnostic coverage and hold out a deterministic $140$-shot test set drawn under a fixed random seed. The remaining $561$ shots are split into a $477$-shot training set and an $84$-shot validation set used for early stopping. Under the shot-level convention used for model evaluation, the train/validation/test splits contain $227/250$, $34/50$, and $72/68$ MARFE-positive/MARFE-negative shots, respectively. No shot is reassigned between partitions across any model variant reported below. The train/validation/test split is used only for downstream predictor training, model selection, and evaluation; all model variants use the same frozen EM-refined reference labels. The training set is shuffled at every epoch and the test set is fixed across all experiments.
Table~\ref{tab:dataset_xtab} cross-classifies the $857$-shot corpus by $\hat{y}$ and the disruption outcome, wherein EM-positive means that at least one frame has $\hat{y}_i=1$. Though it is a MARFE-state convention, an EM-positive shot may be non-disruptive or disrupt for a cause other than MARFE. Only $37/208 \approx 18\%$ of the EM-positive disruptive shots are MARFE-induced in the HL-3 operation log. Such observation also motivates the development of MARFE worsening label by Eq.~\eqref{eq:worsening_label} rather than EM positivity.

\begin{table*}[!t]
\centering
\footnotesize
\caption{Cross-classification of the $857$-shot corpus by the EM-refined label $\hat{y}$ and the disruption outcome. ``MARFE-disr.''~/~``other-disr.'' split disruptive shots by whether the operations log identifies MARFE or another cause as primary.}
\label{tab:dataset_xtab}
\begin{tabular}{@{}lrrrr@{}}
\br
              & MARFE-disr. & other-disr. & non-disr. & Total \\
\mr
EM-positive ($\hat{y}=1$ in some frame)  &  37 & 171 & 177 & 385 \\
EM-negative ($\hat{y}=0$ in every frame) &   7 & 267 & 198 & 472 \\
\mr
Total                                    &  44 & 438 & 375 & 857 \\
\br
\end{tabular}
\end{table*}

\begin{table}[!t]
\centering
\caption{Dataset composition for the frozen train/validation/test split. A shot is \emph{MARFE-positive} if a MARFE/worsening event appears in the evaluated discharge.}
\label{tab:split_composition}
\begin{tabular}{lcccc}
\br
Class & Train & Validation & Test & Total \\
\mr
Total shots & 477 & 84 & 140 & 701 \\
MARFE-positive shots & 227 & 34 & 72 & 333 \\
MARFE-negative shots & 250 & 50 & 68 & 368 \\
\br
\end{tabular}
\end{table}

Model training is performed on a single NVIDIA RTX-5070 GPU. To simulate the real deployment environment and evaluate real-time performance, inference speed tests are conducted on an NVIDIA A100 GPU device. The framework is implemented in PyTorch and optimized with TensorRT for accelerated inference. %Inputs are sampled at $2$\,ms. The prediction target is a $40$\,ms forward worsening label. 
%framework in three steps. First, we check that the physics-gated label refinement improves physical consistency. Second, we assess short-horizon MARFE worsening prediction under a by-shot split to avoid leakage. Third, we run ablations to isolate the contribution of each component.
%We report receiver operating characteristic (ROC) curves and area under the curve (AUC). Because control actions prefer low false-positive rates (FPR), we also discuss behaviour in the low-FPR region of the ROC.  non-overlapping shot bootstrap.

%\subsection{Experimental Setup}

\subsection{Validation of Effectiveness of the Label Refinement Pipeline}\label{sec:label_pipeline_validation}
\begin{figure}[!t]
    \centering
    \includegraphics[width=.85\columnwidth]{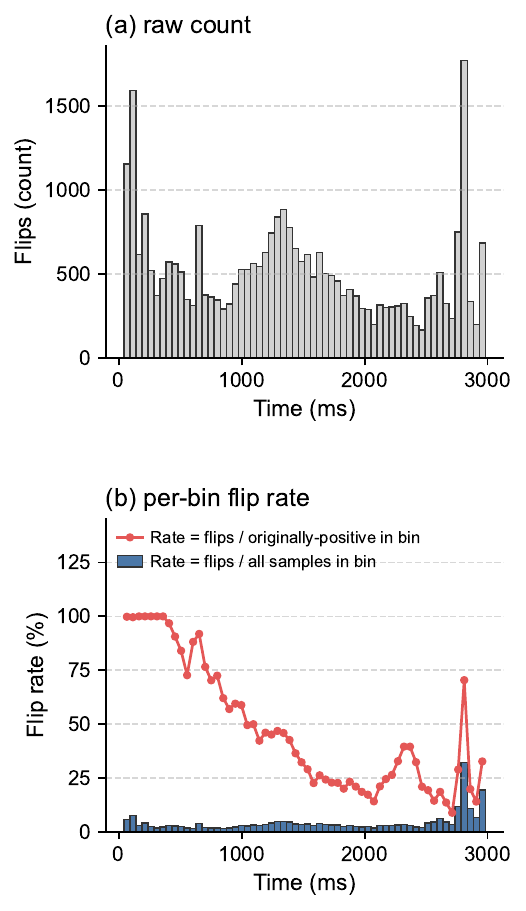}
    \caption{Time distribution of initially positive labels corrected to negative by the refinement algorithm across the entire dataset.}
    \label{fig:correction_dist}
\end{figure}

\begin{figure}[!t]
    \centering
    \includegraphics[width=1.09\columnwidth]{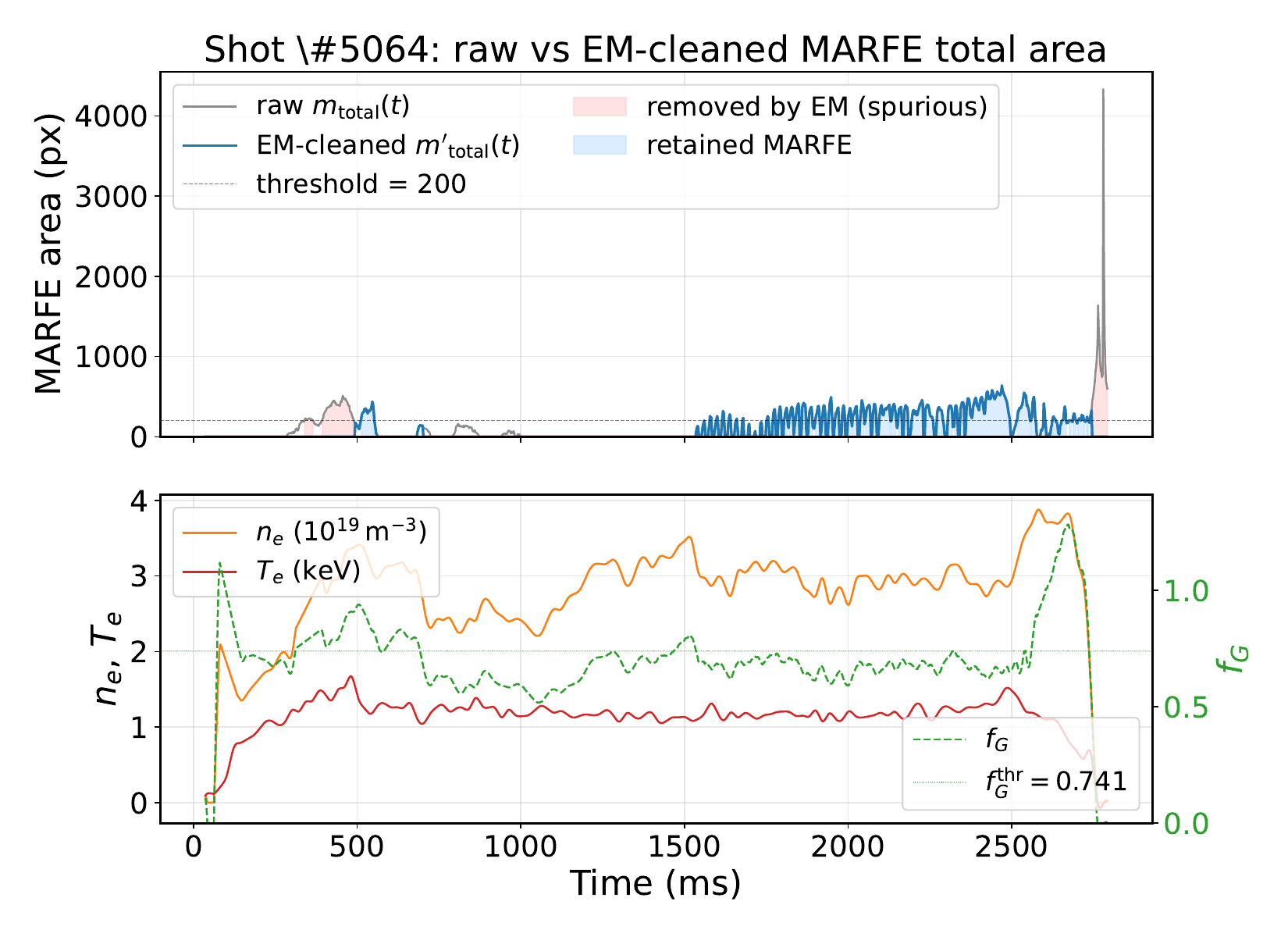}
    \caption{EM-based label refinement on Shot~\#5064 (sub-Greenwald ramp-up case). \emph{Top}: raw $m_\mathrm{total}(t)$, with EM-removed frames in light red and retained frames in light blue. \emph{Bottom}: $n_e$, $T_e$ and Greenwald fraction $f_G$, with the primary physics-score threshold $f_G^\mathrm{thr}=0.741$.}
    \label{fig:em_result_example}
\end{figure}

Due to the lack of a comprehensive manually-labeled ground truth, we validate our label refinement pipeline through several lines of evidence to demonstrate its physical plausibility and downstream effectiveness. Particularly, we verify the effect by analyzing where the algorithm modifies the weak visual labels. Figure~\ref{fig:correction_dist} shows a U-shaped correction pattern: the flip rate is high during early ramp-up and rises again during the late/post-termination phase, while remaining substantially lower during the central flat-top interval, a period known for transient artifacts (wall reflections, divertor bright spots, gas puffs). Figure~\ref{fig:em_result_example} illustrates this behavior on Shot~\#5064: the raw threshold detector triggers on a small number of frames during the early ramp-up phase between roughly $200$ and $700$\,ms, and persistently throughout the late flat-top phase from around $1500$ to $2600$\,ms. The EM refiner removes the early triggers as sub-Greenwald ramp-up artifacts.
Conversely, the EM refiner retains the full late-flat-top window as MARFE-positive. The retention is physically consistent with $f_G$ rising to $\sim\! 1.3$, above the canonical MARFE-onset range $f_G \approx 0.40$--$0.55$ reported across earlier tokamaks~\cite{Lipschultz1984NF,lipschultz1987review}, and $n_e$ reaching $\sim\! 3.5\times 10^{19}\,\mathrm{m}^{-3}$, which together form the canonical density-limit MARFE signature, confirmed by a physicist on visual inspection. A complementary case on Shot~\#5390 (post-disruption afterglow, shown in Figure~\ref{fig:em_result_example_5390} in \ref{app:supp_validation}, illustrates the same EM step correctly handles the second canonical confounder. So the two cases together cover the dominant visual artifacts the refiner is designed to suppress.

A further line of evidence comes from an independent expert audit. We submit a $100$-shot subset to an HL-3 operations physicist for blind binary review. The expert verdict is made independently of the EM output: the expert jointly inspected the raw CCD evolution and the available MARFE-relevant plasma traces, and marked a shot as MARFE-positive only when a physically plausible HFS-bright event is consistent with the operational definition above. Shots lacking such joint visual--physics evidence were marked as non-MARFE. The reviewed subset contains $63$ EM-positive shots spanning both the prior-driven and the GMM-driven retentions of Eq.~\eqref{eq:scoring_function}, and $37$ EM-negative shots spanning the full range of raw visual signal. Table~\ref{tab:expert_audit} reports the shot-level confusion matrix of the initial threshold label $y_\mathrm{init}$ and the EM-refined label $\hat{y}$ against the expert verdict. The EM step lifts precision from $0.45$ to $0.67$ and $F_1$ from $0.62$ to $0.78$, while recall stays above $0.93$: the refined labels retain $42$ of the $45$ shots the expert confirms as real MARFE and remove most of what the expert calls non-MARFE. Notably, this audit evaluates the conditional refinement performance among visually positive candidate shots, but does not measure MARFEs missed by the initial visual detector. A discussion of the dominant residual errors of the present refiner is given in Section~\ref{sec:res_error}.

\begin{table*}[!t]
\centering
\footnotesize
\caption{Initial binary label $y_\mathrm{init}$ versus EM-refined label $\hat{y}$ against an expert-reviewed $100$-shot subset. True positive (TP) and true negative (TN) count shots on which the model and expert agree; false positive (FP) counts model-positive shots the expert reclassifies as non-MARFE; false negative (FN) counts model-removed shots the expert identifies as missed MARFEs. Since all reviewed shots are positive under the shot-level initial visual label ($y_{\mathrm{init}}$), the FN/TN entries of the ($y_{\mathrm{init}}$) row are zero by construction.}
\label{tab:expert_audit}
\begin{tabular}{@{}lcccccccc@{}}
\br
                              & TP & FP & FN & TN & $P$ & $R$ & $F_1$ & Accuracy \\
\mr
Initial $y_\mathrm{init}$     & 45 & 55 &  0 &  0 & 0.450 & 1.000 & 0.621 & 0.450 \\
EM-refined $\hat{y}$          & 42 & 21 &  3 & 34 & 0.667 & 0.933 & 0.778 & 0.760 \\
\br
\end{tabular}
\end{table*}

\begin{table*}[!tbp]
    \centering
    \caption{Performance of the proposed Neural ODE, the Bi-LSTM baseline, the physics-gate ablation, and the information-source ablations on the $140$-shot hold-out test set. Shot-level metrics use the $N=5$ persistence rule at $\theta_\mathrm{act} = 0.5$; ``Lead-med'' is the median label-aligned lead $T^*-t_\mathrm{pred}$ with $P_{25}/P_{75}$ in parentheses. All rows use the same EM-cleaned $b(t)$ target, the same frozen hold-out split, and the same operating threshold $\theta_\mathrm{act}=0.5$, so differences isolate the effect of the model architecture or ablated input-channel group. The no-branch row removes the full $g f_\phi$ physics-gated residual branch and keeps only $f_\theta$.}
    \label{tab:main_results}
    \begin{tabular}{lcccccc}
        \br
        Model & AUC & Sample $F_1$ & Shot $F_1$ & Shot $P$ & Shot $R$ & Lead-med (ms) \\
        \mr
    Physics-Gated Neural ODE (proposed)      & $0.981$          & $\mathbf{0.840}$ & $\mathbf{0.895}$ & $0.901$          & $0.889$          & $+36\,(+14,+248)$ \\
    Bi-LSTM baseline                         & $0.960$          & $0.779$          & $0.883$          & $0.829$          & $\mathbf{0.944}$ & $+61\,(+23,+1244)$ \\
    \mr
    \quad ablation: physics-only ($m'_i{=}0$) & $0.874$          & $0.381$          & $0.746$          & $0.629$          & $0.917$          & $+596\,(+54,+1225)$ \\
    \quad ablation: visual-only (0-D $=0$)    & $\mathbf{0.986}$ & $0.131$          & $0.244$          & $\mathbf{1.000}$ & $0.139$          & $-65\,(-308,-49)$ \\
    \quad ablation: no physics-gated residual branch & $0.958$ & $0.756$ & $0.855$ & $0.782$ & $0.944$ & $+32\,(+12,+52)$ \\
        \br
    \end{tabular}
\end{table*}

\begin{figure}[!t]
    \centering
    \includegraphics[width=\columnwidth]{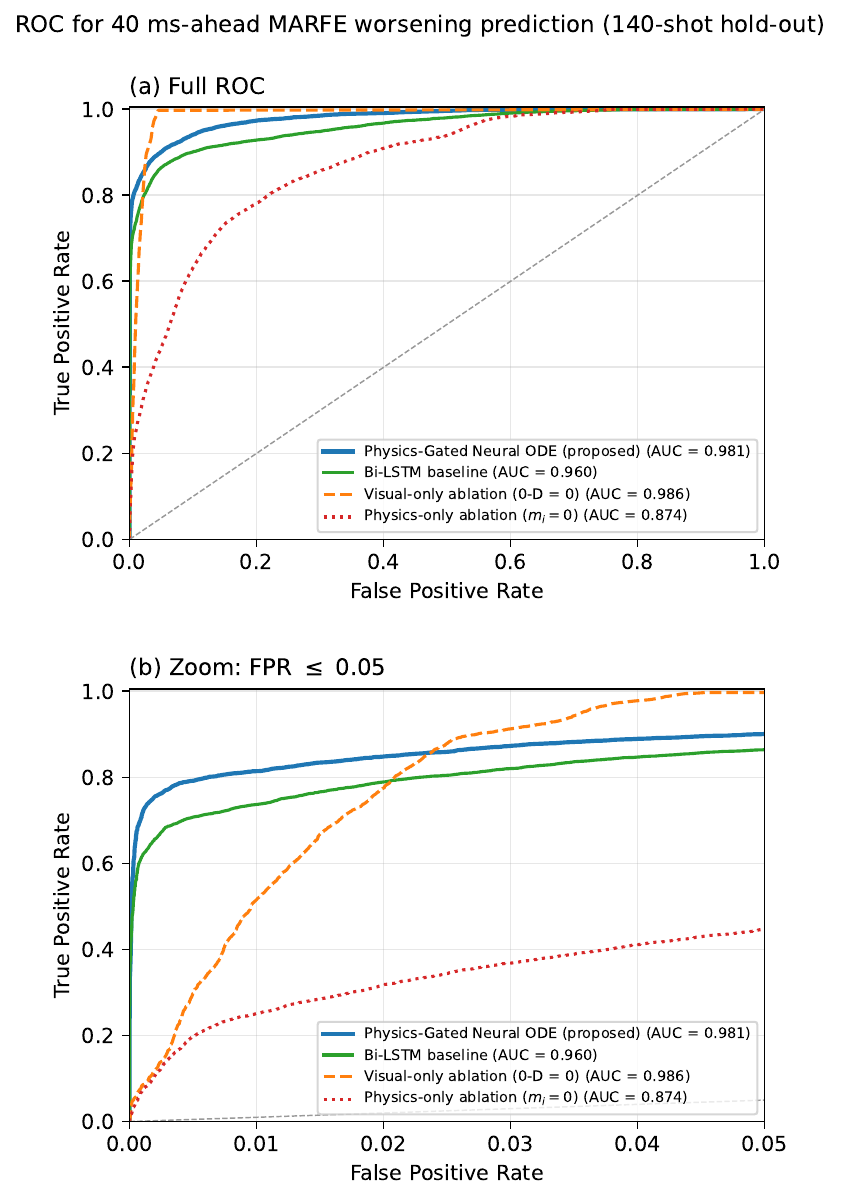}
    \caption{ROC curves for $40$\,ms-ahead MARFE worsening prediction on the $140$-shot test set. Top: full ROC. Bottom: zoom on the low-FPR region (FPR $\le 0.05$).}
    \label{fig:roc_curves}
\end{figure}
\subsection{Performance of the MARFE Worsening Prediction Model}
\subsubsection{Performance Superiority}

With the refined, high-fidelity labels, we train our physics-gated Neural ODE monitor, which produces a per-frame MARFE-intensity probability at the $2$\,ms diagnostic cadence with a $40$\,ms forecasting horizon. For the sequence encoder component of our model, we implement a Bidirectional LSTM (Bi-LSTM) network. For completeness, the internal dynamic functions, $f_{\theta}$ and $f_{\phi}$, are modeled using $2$-layer fully-connected networks, and the final classifier is a $2$-layer fully-connected network with GELU activation and a sigmoid output. To evaluate its performance, we compare it against a Bi-LSTM baseline (the sequence encoder of the proposed model in isolation, without the ODE block) and against two ablations of the full Neural ODE: a \emph{physics-only} variant with the cleaned visual-area channels $m'_U, m'_M, m'_L$ set as zero, and a \emph{visual-only} variant with all $0$-D plasma channels set as zero. All variants are trained from scratch under the same methodology, while the ablations are one-factor-at-a-time, each differing from the full configuration in a single input-channel group.

Therefore, residual imperfections in $\hat{y}_i$ will simultaneously affect all rows, while the physical validity of the target itself has been verified in Table \ref{tab:expert_audit} of Section \ref{sec:label_pipeline_validation}. 
We provide the threshold-independent area under the curve $\mathrm{AUC}$ and the sample-level $F_1$ (per $2$\,ms frame) on the $140$-shot test set in Table~\ref{tab:main_results}. All reported ROC, $F_1$, and shot-level alarm metrics are computed from the total-region output $p(t)=\hat b_{\mathrm{total}}(t)$. The proposed Neural ODE reaches $\mathrm{AUC}=0.981$ and sample-level $F_1=0.840$, while the Bi-LSTM baseline reaches $\mathrm{AUC}=0.960$ and $F_1=0.779$. At the shot level, the Bi-LSTM gives higher recall ($0.944$ vs.\ $0.889$), whereas the Neural ODE produces fewer false positives ($7$ vs.\ $14$), giving higher shot precision ($0.901$ vs.\ $0.829$) and shot-$F_1$ ($0.895$ vs.\ $0.883$). Thus, the Neural ODE provides a better fixed-threshold precision--recall trade-off for closed-loop monitoring.

The ablation results in Table~\ref{tab:main_results} show complementary failure modes. The physics-only model retains high shot recall ($0.917$), indicating that the 0-D plasma trajectory contains genuine precursor information, but it produces many false positives and its sample-$F_1$ falls to $0.381$. The visual-only model has strong threshold-independent ranking ability, with $\mathrm{AUC}=0.986$, but it is poorly calibrated at the fixed operating threshold: its sample-$F_1$ is only $0.131$ and its shot recall is only $0.139$. Removing the physics-gated residual branch reduces the test AUC from $0.981$ to $0.958$ and the sample-level $F_1$ from $0.840$ to $0.756$, while increasing shot-level false positives from $7$ to $19$. The full model combines these two information sources into the most useful fixed-threshold monitor.

Figure~\ref{fig:roc_curves} presents the receiver operating characteristic (ROC) curves and AUC values. Because control actions prefer low false-positive rates (FPR), we also discuss behavior in the low-FPR region of the ROC.  Holding the false-positive rate to $\mathrm{FPR}\le 0.05$, the Neural ODE reaches recall $0.901$, compared with $0.864$ for the Bi-LSTM. Tightening the operating point to $\mathrm{FPR}\le 0.01$, the Neural ODE retains recall $0.814$, compared with $0.737$ for the Bi-LSTM. The maximum sample-level $F_1$ over the threshold sweep is $0.846$ for the Neural ODE and $0.788$ for the Bi-LSTM. This confirms that the Neural ODE remains more favorable in the low-FPR operating region relevant to a closed-loop alarm budget. Because both models receive identical temporal information through the same $40$\,ms sliding window, the gap is attributable to the inductive bias of the continuous-time formulation in the Neural ODE.

\subsubsection{Label-Aligned Lead-Time Analysis}
\label{sec:warning-time-analysis}
\begin{figure}[!tbp]
    \centering
    \includegraphics[width=\columnwidth]{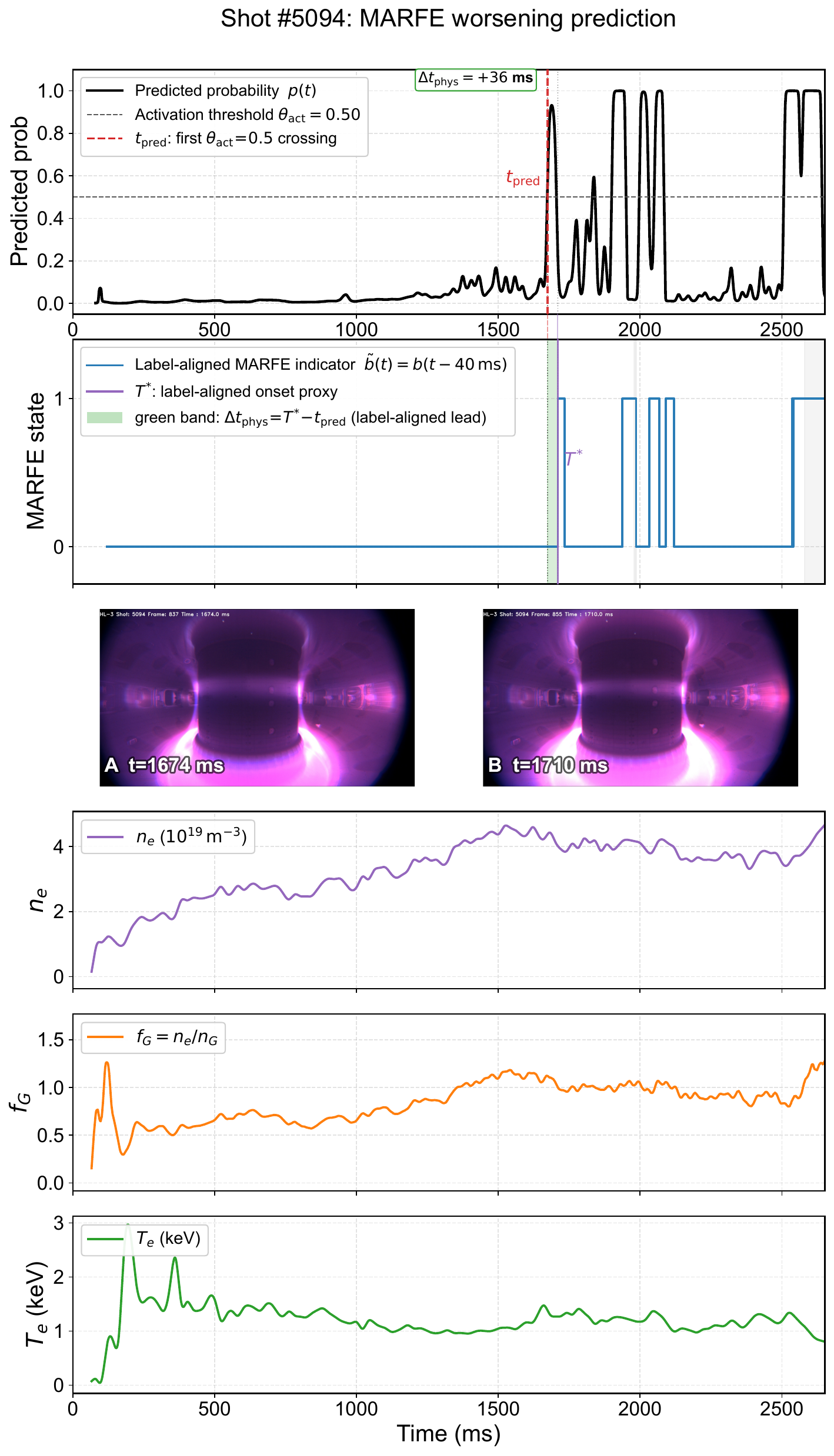}
    \caption{$40$-ms-ahead MARFE prediction on Shot~\#5094 from the frozen test set. \emph{Top}: predicted probability $p(t)$ and the $\theta_\mathrm{act}=0.5$ threshold; the first crossing occurs at approximately $t_\mathrm{pred}=1674$\,ms and is marked by the red dashed line. \emph{Second}: shifted target indicator $\tilde{b}(t)=b(t-40\,\mathrm{ms})$ rising at the label-aligned onset proxy $T^{*}=1710$\,ms (purple); the green band marks the label-aligned lead time $\Delta t_\mathrm{phys}=+36$\,ms. \emph{Middle}: CCD frames before and during MARFE formation. \emph{Bottom}: $n_e$, $f_G$, and $T_e$.}
    \label{fig:shot5094_example}
\end{figure}

Figure~\ref{fig:shot5094_example} illustrates the model's real-time predictive capability on a single shot. Beforehand, to remove possible confusion, we first state the time convention explicitly. The training target $b(t)=1$ in Eq.~\eqref{eq:worsening_label} implies that the MARFE will worsen in the next $40$\,ms. In other words, the training label is shifted $40$\,ms earlier than the label-aligned onset proxy used for evaluation. In Figure~\ref{fig:shot5094_example}, the shifted target indicator $\tilde{b}(t):=b(t-40\,\mathrm{ms})$ rises at the label-aligned onset proxy $T^{*}$. Letting $t_\mathrm{pred}$ denote the time at which the predicted probability first crosses $\theta_\mathrm{act}=0.5$ and $T^*$ denote the label-aligned onset proxy, the lead time is defined as the label-aligned lead $\Delta t_\mathrm{phys}=T^{*}-t_{\text{pred}}$. For Shot~\#5094 from the test set, the predicted probability first crosses the threshold at approximately $t_\mathrm{pred}=1674$\,ms while the label-aligned onset proxy is $T^{*}=1710$\,ms, giving $\Delta t_\mathrm{phys}=+36$\,ms.

\begin{figure}[!t]
    \centering
    \includegraphics[width=\columnwidth]{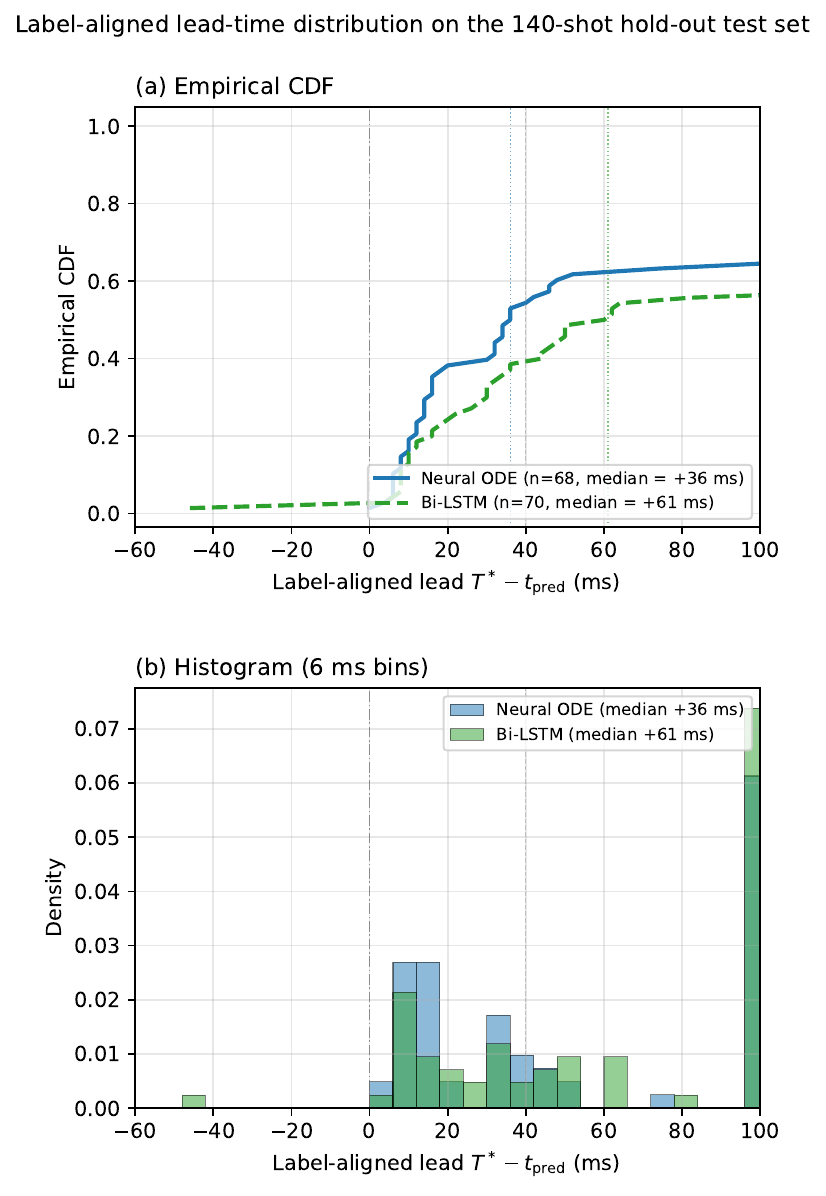}
    \caption{Conditional label-aligned lead-time distribution $T^* - t_{\text{pred}}$ among detected MARFE-positive shots in the $140$-shot frozen test set. Top: empirical CDF over the central plotting window; long positive-tail leads extend beyond the displayed range, so the right edge of the plotted CDF is not forced to one. Bottom: histogram with $6$\,ms bin width; edge bins include values outside the displayed range, while the reported percentiles are computed from the full conditional distribution.}
    \label{fig:warning_cdf}
\end{figure}

Next, we extend the single-shot illustration of Figure~\ref{fig:shot5094_example} to a population-level evaluation over the $140$-shot test set. For event-timing analysis, $216$ MARFE events from $124$ shots were annotated from the raw, pre-EM regional bright-area traces using the same $200$-pixel total-area criterion as Section~2.2.1. A valid event contains at least five consecutive positive frames ($\ge 10$\,ms), and two events are separated by at least $80$\,ms of quiescence. Events extending beyond the manually recorded disruption time, or beginning within $5$\,ms of it, were excluded as post-disruption afterglow. For multi-event shots, the shot-level confusion matrix is computed per discharge: a positive declaration requires one persistent alarm within the union of MARFE/worsening intervals, whereas the warning-time statistic uses the first label-aligned onset proxy as the first-useful-alarm reference. For each MARFE-positive shot, we take the label-aligned onset proxy as $T^{*} = t_\mathrm{gt} + 40$\,ms, where $t_\mathrm{gt}$ is the first frame at which the forward-looking EM-cleaned label $b(t)$ crosses $0.5$; the $40$\,ms offset reflects the forward-look horizon of Eq.~\eqref{eq:worsening_label} ($b(t) = \tilde{b}(t+40\,\mathrm{ms})$, so the rising edge of $b$ precedes the shifted label indicator $\tilde{b}$ by exactly $40$\,ms). The label-aligned lead $\Delta t_\mathrm{phys} = T^{*} - t_\mathrm{pred}$ (positive values mean the alarm precedes the label-aligned onset proxy) is computed on MARFE-positive test shots for which the predicted probability crosses $\theta_\mathrm{act}=0.5$ at some point in the discharge; shots, on which the model never activates, are counted as shot-level false negatives and are excluded from the warning-time distribution. Figure~\ref{fig:warning_cdf} overlays the central plotting window of the cumulative distribution functions of $\Delta t_\mathrm{phys}$ for the Neural ODE and the Bi-LSTM baseline on this test set; long positive-tail leads that extend beyond the displayed range are still retained in the percentile statistics. The Neural ODE distribution is centered at a median of $+36$\,ms with interquartile range $(+14,+248)$\,ms, close to the $40$-ms design horizon. The Bi-LSTM has a larger median lead of $+61$\,ms with IQR $(+23,+1244)$\,ms, but this earlier triggering is accompanied by more false positives ($14$ versus $7$) and a heavier early-trigger tail.

%Additionally, we report a per-event recall as a sanity check in Table \ref{tab:main_results}. On the $33$ manually-annotated test events, the Neural ODE produces a sustained alarm within $100$\,ms before $T^{*}$ on $63.6\%$ of events with a median captured-event lead of $+34$\,ms; the Bi-LSTM and the visual-only ablation track this rate within $ \pm 2$ events, indicating that the recall ceiling at this annotated sample size is set by the task itself rather than by the architectural choice. The physics-only ablation behaves differently: it reaches only $39.4\%$ at $W = 100$\,ms but its median captured-event lead is $+62$\,ms and its recall catches up to the visual-feature models at $W \ge 500$\,ms, the quantitative signature of a long-but-imprecise predictor whose $0$-D plasma trajectory trends towards MARFE conditions earlier than the visual area starts to ramp. The two channels are therefore complementary by design, motivating the two-channel formulation of Eq.~(\ref{eq:ode}). %Extending the input to include equilibrium-reconstructed position channels is the natural extension and is left for future work (Section~\ref{sec:discussion}). For the closed-loop monitor deployment target, the central operational metric remains the frame-level discriminant quality reported in Table~\ref{tab:main_results} and Figure~\ref{fig:roc_curves}.

\subsection{Real-Time Performance Evaluation and System Deployment}
\begin{figure}[!t]
\centering
\includegraphics[width=\columnwidth]{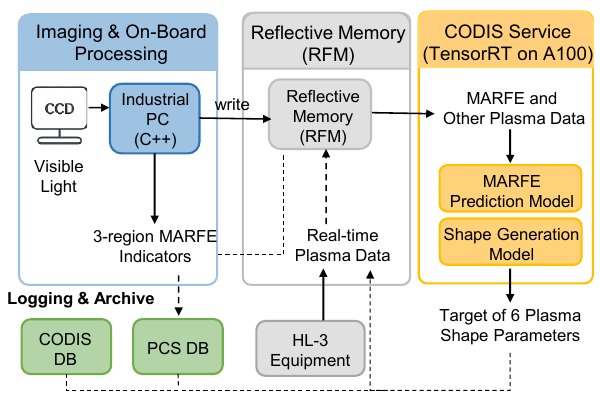}
\caption{Real-time deployment architecture: data flow from the CCD camera through the industrial PC (feature extraction and fixed-parameter EM/GMM posterior inference) to the CODIS server (TensorRT Neural-ODE inference and control-target generation) and the Plasma Control System.}
\label{fig:deployment}
\end{figure}

A critical requirement for any disruption warning system is its ability to operate within the stringent time constraints of a plasma control loop. To validate the real-time capability of our framework, we deploy the real-time feature extractor, the fixed-parameter EM/GMM posterior-inference step, and the trained Neural ODE monitor. Figure \ref{fig:deployment} provides the high-throughput, low-latency processing design for deployment.

The data acquisition and processing pipeline begins with a CCD camera that captures visible-light frames in real time. An industrial PC, located near the diagnostic, performs initial image processing (masking, thresholding, denoising) and fixed-parameter EM/GMM posterior inference using a C++ implementation to generate cleaned regional MARFE-area features. These features are written to a reflective memory segment at a $1$-ms update period. Concurrently, a TensorRT-optimized inference service, running on a server-grade machine, polls this memory at $0.1$-ms intervals. Upon receiving new feature data, the service executes the physics-gated Neural ODE model to compute the MARFE worsening probabilities. Subsequently, it calculates six configuration targets for plasma shape control (major radius $R$, vertical position $Z$, elongation $\kappa$, upper and lower triangularities $\delta_{u}, \delta_{l}$, and strike point/gap geometry).

To rigorously quantify performance, we benchmark the execution time of each pipeline stage. As summarized in Table \ref{tab:performance}, the initial image processing, fixed-parameter EM/GMM posterior inference, and data transfer costs an average of $0.7$\,ms. The core model inference, which initially takes approximately $1.0$\,ms in a standard PyTorch environment on an NVIDIA A100 GPU, is significantly accelerated to just $0.3$\,ms after TensorRT optimization. %This optimization is crucial for minimizing the total loop latency. 
The entire pipeline, from image acquisition to the generation of control targets, achieves an average cadence of $1$\,ms. This performance is well within the requirements for real-time feedback control, enabling the system to supervise plasma shape proactively to steer the discharge away from MARFE-prone conditions while maintaining operational performance. Both the raw optical MARFE indicators and the final inferred probabilities are archived in the CODIS and Plasma Control System (PCS) databases for post-shot analysis.

\begin{table}[!t]
\centering
\small % Or \footnotesize, or \scriptsize
\caption{Component-wise timing analysis of the real-time MARFE prediction pipeline.}
\label{tab:performance}
\begin{tabular}{lc}
\hline
\textbf{Pipeline Stage} & \textbf{Average Time (ms)} \\
\hline
Image Preprocessing + EM & $0.6$ \\
Model Inference (PyTorch) & $\sim1.0$ \\
Model Inference (TensorRT) & $\mathbf{0.3}$ \\
Control Target Calculation & $<0.1$ \\
\hline
\textbf{Total End-to-End Latency} & $\sim\mathbf{1.0}$ \\
\hline
\end{tabular}
\end{table}

\section{Discussions}\label{sec:discussion}

\subsection{Distinction between Physics-Gated and Physics-Informed}
\label{sec:phy_gate}
We adopt the term \emph{physics-gated} to distinguish this construction from the more common physics-informed neural network (PINN) of Raissi et al.~\cite{Raissi2019JCP}, which incorporates a partial differential equation residual into the loss. Here the physical prior enters at two architectural levels instead: (i) the EM physics score $s_i$ of Section~\ref{sec:physics score} encodes 
Greenwald-fraction and electron-cooling criteria as a sample-specific Bayesian prior on the label, and (ii) the sigmoid gate $g(f_G, T_e)$ embeds the MARFE-onset threshold as a soft architectural constraint on the ODE dynamics.

\subsection{Analysis of Dominant Residual Errors in Label Refinement}
\label{sec:res_error}

The expert audit of Section~\ref{sec:label_pipeline_validation} pinpoints the dominant residual error of the present four-channel formulation as two position-induced visual confounders. The $21$ false positives are genuinely high-density discharges that drift inward and brighten on the HFS limiter. The density clears the $f_G \ge 0.40$ gate, but the CCD brightness comes from limiter contact rather than from a radiative thermal instability. Another $3$ false negatives reflect the opposite mechanism: the plasma drifts out of the Thomson sightline during a real high-density MARFE-prone discharge, the measured $n_e$ drops spuriously, and the under-scored physics prior wrongly removes a true MARFE. Both errors hinge on plasma position relative to the limiter and the diagnostic line of sight rather than on the local $(n_e, T_e, f_G)$ state. Therefore, extending the input with the equilibrium-reconstructed channels $(R_\mathrm{geom}, Z_\mathrm{geom}, a_\mathrm{out}, \kappa)$ is the natural next step, with no architectural change beyond the input dimension.

\subsection{Cross-Device Transferability}

The deployment of this monitor is portable to other medium- to large-size tokamaks under modest diagnostic requirements: a line-averaged or core $n_e$ (interferometer or Thomson), a core $T_e$ signal (Thomson on HL-3, or an ECE/soft-X-ray proxy on other devices), the plasma current $I_p$ and minor radius $a$ (for the Greenwald-fraction calculation), and 
a visible-light CCD that resolves at least $10$ pixels poloidally in the HFS region. Quantities that require recalibration on a new device are the primary thresholds in Table~\ref{tab:thresholds} (re-derived via Youden's $J$ on a small labelled bootstrap of $50$--$100$ shots of the new device), the early-time hard-gate cutoff $t_\mathrm{cut}$ (defaulting to $300$\,ms on HL-3 but device-dependent through the breakdown phase), and the forecasting horizon. The density gate based on $f_G$ is transferable at the level of procedure, but its numerical threshold should be recalibrated on each device because the available real-time density proxy may differ from the conventional line-averaged density, whereas the time gate $t<t_{\mathrm{cut}}$ should be re-calibrated, replaced by a normalized shot-phase variable, or omitted according to the target device and deployment scenario.
 ITER, EAST and JET are the most natural targets for cross-device validation, where data continuity from longer pulses additionally motivates the continuous-time monitor formulation adopted here. As HL-3 currently relies on gas puffing as the dominant fuelling pathway, the present framework is trained and validated on gas-fuelled discharges; extending it to pellet-fuelled discharges is a natural direction once HL-3 implements pellet injection capability.

\section{Conclusion and Future Research}\label{sec:conclusion}

This work has developed a physics-gated, continuous MARFE monitor for the HL-3 tokamak, and made two principal contributions. First, we have proposed a physics-scored, weighted EM pipeline that refines noisy visual labels using $(n_e, T_e, f_G, t)$ to construct an empirical physics prior, recovering high-fidelity training labels from imperfect camera data. Second, we have designed a continuous-time, physics-gated Neural ODE backbone whose dynamics are modulated by 
a sigmoid gate on $(f_G, T_e)$, producing a stable per-frame MARFE-intensity discriminant suitable for closed-loop control feedback. 
On a $140$-shot test set, the proposed framework attains $\mathrm{AUC}=0.981$ and sample-level $F_1=0.840$, against $\mathrm{AUC}=0.960$ and $F_1=0.779$ for the matched-protocol Bi-LSTM baseline. At the shot level, it reduces false positives from $14$ to $7$ and improves shot precision from $0.829$ to $0.901$, while the Bi-LSTM retains higher shot recall. Meanwhile, information-source ablations confirm that the visual and $0$-D channels carry complementary signals for fixed-threshold monitoring. The deployed feature-extraction, fixed-parameter EM/GMM posterior-inference, and Neural-ODE inference pipeline runs within the $1$-ms cycle time.

In the future, we plan to conduct closed-loop experiments, feeding the generated shape targets to the HL-3 plasma control system to actively suppress MARFEs.
%This work represents a significant step toward the intelligent, real-time control systems essential for the safe and steady operation of future fusion reactors like ITER.

\section*{Acknowledgments}
The authors would like to thank the entire HL-3 team for providing experimental data. This work was supported in part by the National Key Research and Development Program of China under Grant 2024YFE03020001.

%\onecolumn
\appendix
\section{Supplementary Validation of the Label Refinement Pipeline}\label{app:supp_validation}

We provide three complementary pieces of evidence for the EM step beyond the main-body validation of Section~\ref{sec:label_pipeline_validation}.

\noindent\textbf{A.1 Bucket-level audit of the EM-derived flips.} We trace where the EM-derived flips come from by classifying the $33{,}672$ frame-level positive-to-negative flips into four mutually exclusive buckets according to the decisive factor of the Bayesian posterior on each frame (Table~\ref{tab:em_audit}). Buckets A and B correspond to samples on which the hard gates ($t < 300$\,ms for A; $f_G < 0.40$ at $t \ge 300$\,ms for B) fix $s_i = 0.01$, so the flip is in effect produced by the gate itself and its validity reduces to the validity of the physics rule. Bucket C captures samples with $s_i \le 0.10$ in the absence of positive physical indicator, so the prior alone drives the flip. Bucket D collects samples that pass both hard gates and have $s_i > 0.10$, where the data-driven GMM posterior is what catches confounders such as divertor brightening or late-shot afterglow. Buckets A--C together account for $67.8\%$ of the flips and are directly traceable to literature-grounded physics rules, while bucket D supplies the remaining $32.2\%$ that the learned Gaussian mixture flips beyond what the rules alone would catch. Removing the GMM and keeping only the two hard gates would inflate the EM-positive shot count on the full $857$-shot dataset from $385$ to $485$, an excess of $26\%$ in the full-corpus EM-positive shot count, confirming that the two components are operationally complementary.

\begin{table*}[!ht]
\centering
\footnotesize
\caption{Audit of the $33{,}672$ frame-level flips ($y_\mathrm{init}=1 \to \hat{y}_i=0$) produced by the EM run, grouped by which factor of the Bayesian posterior is decisive on each frame.}
\label{tab:em_audit}
\renewcommand{\arraystretch}{1.2}
\begin{tabular}{@{}p{0.30\textwidth}rrp{0.45\textwidth}@{}}
\br
\textbf{Trigger} & \textbf{Count} & \textbf{Share} & \textbf{Physical interpretation} \\
\mr
(A) Early-time gate, $t < 300$\,ms                          & 4{,}639           & $13.8\%$        & Breakdown / early ramp-up; gas-puff and wall-reflection artifacts. \\
(B) Density gate, $f_G < 0.40$ at $t \ge 300$\,ms           & 9{,}139           & $27.1\%$        & Extended ramp-up; sub-Greenwald regime where a MARFE is physically improbable. \\
(C) Low $s_i \le 0.10$, no gate activated                       & 9{,}067           & $26.9\%$        & No positive physical indicator; low-density or divertor-flare artifacts. \\
(D) GMM posterior, $s_i > 0.10$, no gate activated              & 10{,}827          & $32.2\%$        & Residual ambiguous regime resolved by the data-driven Gaussians. \\
\mr
\textbf{(A+B+C) Physics-rule-driven}                        & \textbf{22{,}845} & \textbf{67.8\%} & Traceable to literature-grounded physics rules. \\
\textbf{(D) EM-posterior-driven}                            & \textbf{10{,}827} & \textbf{32.2\%} & Produced by the data-driven Gaussian mixture only. \\
\br
\end{tabular}
\end{table*}

\begin{figure*}[!t]
    \centering
    \includegraphics[width=2\columnwidth]{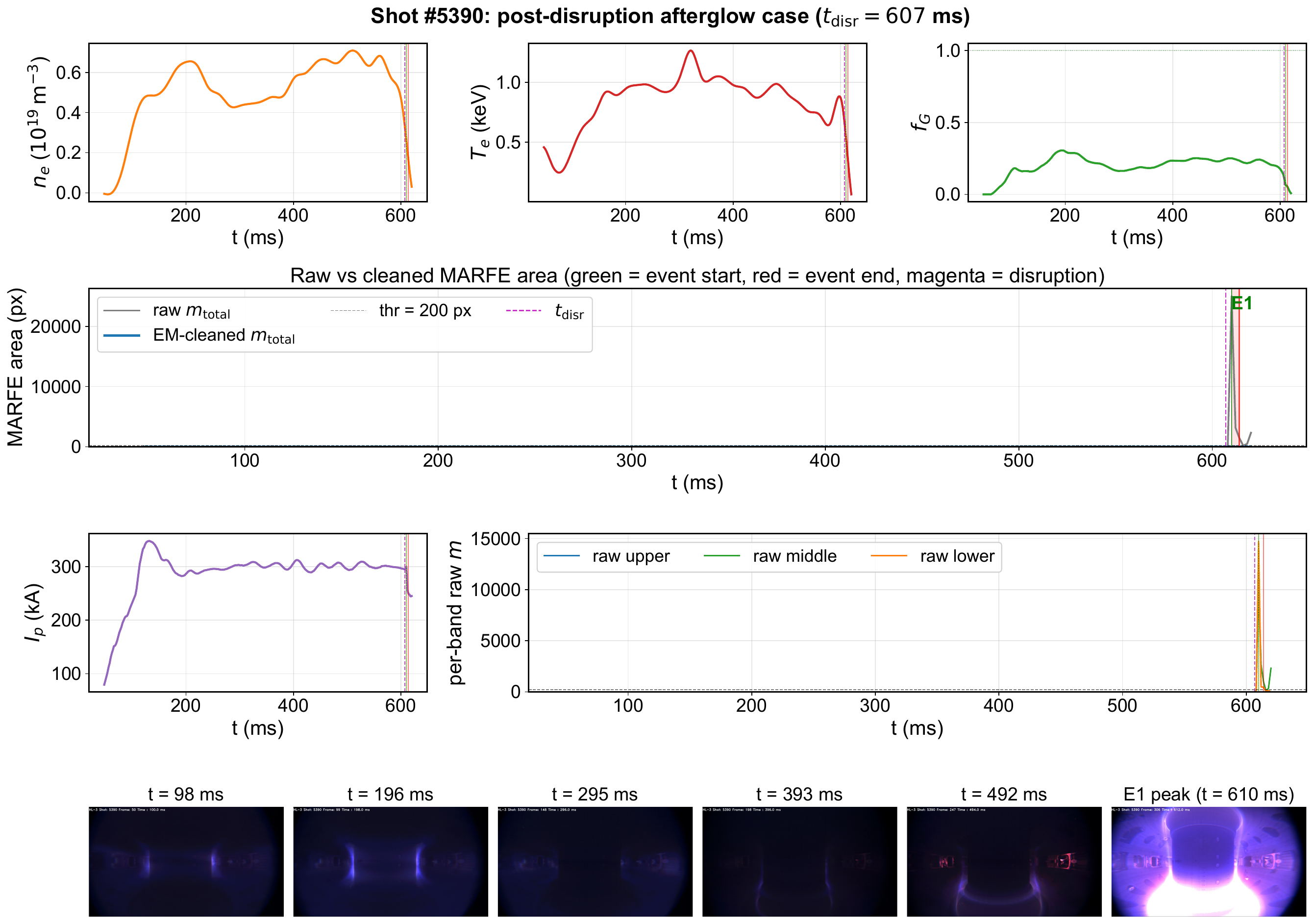}
    \caption{Complementary EM-based label refinement on Shot~\#5390, an early-disruption shot with a post-current-quench afterglow artifact ($t_\mathrm{disr}=607$ ms). The post-current-quench afterglow brightens the entire CCD field of view (raw $m_\mathrm{total}$ surges in the second row), but the EM refiner correctly relabels the affected frames as non-MARFE, driving the cleaned $m_\mathrm{total}$ to zero, on the basis of the simultaneous collapse of $n_e$ and $f_G$ in the top row.}
    \label{fig:em_result_example_5390}
\end{figure*}

\noindent\textbf{A.2 Second case study: post-disruption afterglow on Shot~\#5390.} Figure~\ref{fig:em_result_example_5390} illustrates the EM step on the complementary visual confounder: a post-current-quench afterglow that brightens the entire CCD field of view, in contrast with the sub-Greenwald ramp-up case of Shot~\#5064 in the main body (Figure~\ref{fig:em_result_example}). The raw $m_\mathrm{total}$ surges after the disruption time $t_\mathrm{disr}=607$\,ms, but the EM refiner correctly relabels the affected frames as non-MARFE on the basis of the simultaneous collapse of $n_e$ and $f_G$. Together the two cases cover the dominant visual artifacts the refiner is designed to suppress.

\noindent\textbf{A.3 Monte-Carlo sensitivity of the physics score under diagnostic noise.} We quantify how upstream diagnostic noise propagates to the physics-score component of the EM refiner with a Monte-Carlo study on the full $857$-shot, $888{,}529$-frame corpus, by intentionally adding noise to the Thomson-derived core-density signal $n_e$ and core-temperature signal $T_e$. Afterward, we recompute the physics score $s_i$ from Eq.~\eqref{eq:scoring_function}, and count threshold-crossing changes of the score decision at $s_i=0.5$ over $M=200$ realizations. Table~\ref{tab:mc_noise} reports the corresponding score-decision flip rate. At the nominal noise level, the overall sample-level score-decision flip rate is $1.66\%$, with the borderline-score stratum elevated to $3.03\%$ as expected. Even under more aggressive noise, the overall rate stays under $3\%$, indicating that the physics-score component of the EM refiner is stable under the diagnostic noise budget.

\begin{table*}[!ht]
\centering
\footnotesize
\caption{Monte-Carlo sensitivity of the physics-score decision under multiplicative Gaussian noise on $(n_e, T_e)$, reported as mean $\pm$ s.d.\ over $M=200$ realizations on the $857$-shot, $888{,}529$-frame corpus. \emph{Borderline}: $0.10 < s_i < 0.70$.}
\label{tab:mc_noise}
\begin{tabular}{@{}lccccc@{}}
\br
Noise regime & $\sigma_{n_e}$ & $\sigma_{T_e}$ & Overall flip (\%) & Extreme (\%) & Borderline (\%) \\
\mr
Conservative & $3\%$  & $7\%$  & $1.108 \pm 0.009$ & $0.715 \pm 0.009$ & $1.941 \pm 0.023$ \\
Nominal      & $5\%$  & $10\%$ & $1.658 \pm 0.010$ & $1.013 \pm 0.010$ & $3.027 \pm 0.028$ \\
Aggressive   & $10\%$ & $15\%$ & $2.763 \pm 0.014$ & $1.578 \pm 0.012$ & $5.276 \pm 0.036$ \\
\br
\end{tabular}
\end{table*}

\section{Notations and Abbreviations}
\label{app:notation}
Please refer to Table~\ref{tab:notations_appendix} and Table \ref{tab:abbrev}.
\begin{table}
%\small
\caption{Nomenclature and List of Symbols.}\label{tab:notations_appendix}
\centering
\begin{tabular}{p{0.11\textwidth} p{0.325\textwidth}}
\toprule
\textbf{Symbol} & \textbf{Description} \\
\midrule
% \endfirsthead

% \multicolumn{2}{c}{{\tablename\ \thetable{} -- continued from previous page}}\\
% \br
% \textbf{Symbol} & \textbf{Description} \\
% \mr
% \endhead

% \multicolumn{2}{r}{{Continued on next page}}\\
% \mr
% \endfoot

% \br
% \endlastfoot

\multicolumn{2}{l}{\underline{General Plasma Parameters}}\\
$n_e$ & Core electron density.\\
$f_G$ & Normalized density ($n_e/n_g$). \\
$T_e$ & Core electron temperature.\\
$q_{95}$ & Safety factor at the 95\% flux surface.\\
$I_p$ & Plasma current.\\
$R, a$ & Plasma major and minor radius, respectively.\\
$\kappa, \delta$ & Plasma elongation and triangularity, respectively.\\
$Z$ & Vertical position of the plasma column.\\
$l_i$ & Plasma internal inductance.\\
$P_{\text{NBI}}$, $P_{\text{ECRH}}$, $P_{\text{LHCD}}$ & Heating power from NBI, ECRH, and LHCD systems.\\
%\addlinespace

\multicolumn{2}{l}{\underline{Image-Derived Features}}\\
$m_U, m_M, m_L$ & Initial MARFE area features from upper, middle, and lower camera zones.\\
$m'_{U}, m'_{M}, m'_{L}$ & Cleaned (refined) MARFE area features. \\% used as model input.\\
$y_{\text{init}}$ & Initial binary MARFE label. \\% derived from visual features before refinement.\\
%$M_i(t)$ & Time series of the cleaned features for region $i$.\\
%\addlinespace

\multicolumn{2}{l}{\underline{Physics-Scoring and Label Refinement (EM Algorithm)}}\\
$s_i$ & Physics consistency score for sample $i$, used as a prior.\\% in the EM algorithm.\\
$x_i$ & Feature vector $[n_e, T_e, f_{G}, t]_i$. \\% used in the EM algorithm.\\
$z_i$ & Latent variable representing the true MARFE state for sample $i$.\\
$\hat{y}_i$ & Refined binary MARFE label after EM. \\% the EM algorithm.\\
$\gamma(z_{i,\text{pos}})$ & Posterior probability that sample $i$ is a true MARFE.\\
$\theta$ & Set of parameters $\{\mu, \Sigma\}$ for GMM. \\%the Gaussian Mixture Model (GMM).\\
$\mu_{\text{pos}}, \Sigma_{\text{pos}}$ & Mean and covariance of the positive GMM component. \\% for the MARFE state.\\
$\mu_{\text{neg}}, \Sigma_{\text{neg}}$ & Mean and covariance of the negative GMM component. \\% for the non-MARFE state.\\
$\alpha$ & Learning rate for the tempered update in the M-step.\\
%\addlinespace

\multicolumn{2}{l}{\underline{Physics-Gated Neural ODE}}\\
$b(t)$ & The ground-truth ``MARFE worsening'' label for $t$.\\
$\tilde{b}(t)$ & Label-aligned MARFE indicator shifted to the operational onset proxy, $\tilde{b}(t)=b(t-40\,\mathrm{ms})$.\\
$\mathbf{A}(t)$ & Input sequence over a $40$ ms window until $t$.\\
%$h_0$ & Initial latent state by the sequence encoder.\\
$\mathbf{h}(t)$ & Latent state evolved by the Neural ODE.\\
$f_\theta$ & Neural network defining $\mathrm{d}h/\mathrm{d}t$.\\
$g(f_G, T_e)$ & Physics-guided gate modulating the ODE dynamics.\\
$f_\phi$ & Learned physics-conditioned residual dynamics, modulated by $g$.\\
$\mathcal{L}$ & Total uncertainty-weighted binary cross-entropy loss.\\
$\sigma$ & Learnable uncertainty for each prediction task.\\
$\Delta t_{\text{phys}}$ & Lead time ahead of the worsening event.\\
\bottomrule
\end{tabular}
\end{table}
%\twocolumn
%\clearpage
% \subsection*{A. Patch Merging}

% \subsection*{B. Self-Attention Block}

\begin{table}[!t]
\caption{List of abbreviations.}\label{tab:abbrev}
\centering
\begin{tabular}{p{0.13\textwidth} p{0.32\textwidth}}
\toprule
\textbf{Abbreviation} & \textbf{Definition} \\
\midrule
\multicolumn{2}{l}{\underline{Physics and devices}}\\
MARFE   & Multifaceted Asymmetric Radiation From the Edge.\\
HFS     & High-Field Side.\\
HL-3    & HL-3 tokamak (SWIP, China).\\
ITER    & International Thermonuclear Experimental Reactor.\\
EAST    & Experimental Advanced Superconducting Tokamak.\\
JET     & Joint European Torus.\\
SWIP    & Southwestern Institute of Physics.\\

\multicolumn{2}{l}{\underline{Diagnostics and heating}}\\
CCD     & Charge-Coupled Device.\\
ECE     & Electron Cyclotron Emission.\\
HCN     & Hydrogen Cyanide (interferometer).\\
FIR     & Far-Infrared (interferometer).\\
HCD     & Heating and Current Drive.\\
NBI     & Neutral Beam Injection.\\
ECRH    & Electron Cyclotron Resonance Heating.\\
LHCD    & Lower Hybrid Current Drive.\\

\multicolumn{2}{l}{\underline{Real-time deployment}}\\
PCS     & Plasma Control System.\\
CODIS   & Central Online Data Integration System (HL-3).\\
GPU     & Graphics Processing Unit.\\

\multicolumn{2}{l}{\underline{Machine learning}}\\
EM      & Expectation--Maximization.\\
GMM     & Gaussian Mixture Model.\\
ODE     & Ordinary Differential Equation.\\
PDE     & Partial Differential Equation.\\
PINN    & Physics-Informed Neural Network.\\
RNN     & Recurrent Neural Network.\\
LSTM    & Long Short-Term Memory.\\
Bi-LSTM & Bidirectional Long Short-Term Memory.\\
GELU    & Gaussian Error Linear Unit.\\
RK4     & Fourth-order Runge--Kutta integration scheme.\\
BCE     & Binary Cross-Entropy.\\
ROI     & Region of Interest.\\

\multicolumn{2}{l}{\underline{Evaluation metrics}}\\
ROC     & Receiver Operating Characteristic.\\
AUC     & Area Under the ROC Curve.\\
CDF     & Cumulative Distribution Function.\\
IQR     & Interquartile Range.\\
RMS     & Root-Mean-Square.\\
TP, FP  & True Positive, False Positive.\\
TN, FN  & True Negative, False Negative.\\
TPR     & True Positive Rate (Recall, Sensitivity).\\
FPR     & False Positive Rate.\\
\bottomrule
\end{tabular}
\end{table}

\section*{References}

\bibliographystyle{unsrt}
\bibliography{references} 

@article{hender2007mhd,
  title={{MHD} stability, operational limits and disruptions},
  author={Hender, TC and Wesley, JC and Bialek, J and Bondeson, A and Boozer, AH and Buttery, RJ and Garofalo, A and Goodman, TP and Granetz, RS and Gribov, Y and others},
  journal={Nuclear Fusion},
  volume={47},
  number={6},
  pages={S128},
  year={2007},
  publisher={IOP Publishing}
}

@book{butcher2016numerical,
  title={Numerical methods for ordinary differential equations},
  author={Butcher, John Charles},
  year={2016},
  publisher={John Wiley \& Sons}
}

@article{aymar2002iter,
  title={The {ITER} design},
  author={Aymar, R and Barabaschi, P and Shimomura, Y},
  journal={Plasma Physics and Controlled Fusion},
  volume={44},
  number={5},
  pages={519},
  year={2002},
  publisher={IOP Publishing}
}

@article{bigot2019iter,
  title={ITER construction and manufacturing progress toward first plasma},
  author={Bigot, Bernard},
  journal={Fusion Engineering and Design},
  volume={146},
  pages={124--129},
  year={2019},
  publisher={Elsevier}
}

@article{huber2007improved,
  title={Improved radiation measurements on {JET}--First results from an upgraded bolometer system},
  author={Huber, A and McCormick, K and Andrew, P and de Baar, MR and Beaumont, P and Dalley, S and Fink, J and Fuchs, JC and Fullard, K and Fundamenski, W and others},
  journal={Journal of Nuclear Materials},
  volume={363},
  pages={365--370},
  year={2007},
  publisher={Elsevier}
}

@article{murari2024control,
  title={A control oriented strategy of disruption prediction to avoid the configuration collapse of tokamak reactors},
  author={Murari, Andrea and Rossi, Riccardo and Craciunescu, Teddy and Vega, Jes{\'u}s and Gelfusa, Michela},
  journal={Nature Communications},
  volume={15},
  number={1},
  pages={2424},
  year={2024},
  publisher={Nature Publishing Group UK London}
}

@article{stuart2021petra,
  title={PETRA: A generalised real-time event detection platform at {JET} for disruption prediction, avoidance and mitigation},
  author={Stuart, CI and Artaserse, G and Card, P and Carvalho, IS and Felton, R and Gerasimov, SN and Goodyear, A and Henriques, RB and Karkinsky, D and Lomas, PJ and others},
  journal={Fusion Engineering and Design},
  volume={168},
  pages={112412},
  year={2021},
  publisher={Elsevier}
}

@article{lipschultz1987review,
  title={Review of {MARFE} phenomena in tokamaks},
  author={Lipschultz, Bruce},
  journal={Journal of Nuclear Materials},
  volume={145},
  pages={15--25},
  year={1987},
  publisher={Elsevier}
}

@article{Lipschultz1984NF,
  title={{MARFE}: An edge plasma phenomenon},
  author={Lipschultz, B and LaBombard, B and Marmar, ES and Pickrell, MM and Terry, JL and Watterson, R and Wolfe, SM},
  journal={Nuclear Fusion},
  volume={24},
  number={8},
  pages={977},
  year={1984},
  publisher={IOP Publishing}
}

@article{Greenwald1988NF,
  title={A new look at density limits in tokamaks},
  author={Greenwald, Martin and Terry, JL and Wolfe, SM and Ejima, S and Bell, MG and Kaye, SM and Neilson, GH},
  journal={Nuclear Fusion},
  volume={28},
  number={12},
  pages={2199},
  year={1988},
  publisher={IOP Publishing}
}

@article{Kelly2001PoP,
  title={Thermal instability theory analysis of multifaceted asymmetric radiation from the edge ({MARFE}) in Tokamak Experiment for Technology Oriented Research ({TEXTOR})},
  author={Kelly, FA and Stacey, WM and Rapp, J and Brix, M},
  journal={Physics of Plasmas},
  volume={8},
  number={7},
  pages={3382--3390},
  year={2001},
  publisher={American Institute of Physics}
}

@article{Samm1999JNM,
  author  = {U. Samm and M. Brix and F. Durodi{\'e} and M. Lehnen and A. Pospieszczyk and J. Rapp and G. Sergienko and B. Schweer and M. Z. Tokar and B. Unterberg},
  title   = {{MARFE} feedback experiments on {TEXTOR-94}},
  journal = {Journal of Nuclear Materials},
  year    = {1999},
  volume  = {266-269},
  pages   = {666--672},
  doi     = {10.1016/S0022-3115(98)00516-9}
}

@article{Murari2010TPS,
  title={Algorithms for the automatic identification of {MARFEs} and {UFOs} in {JET} database of visible camera videos},
  author={Murari, Andrea and Camplani, Massimo and Cannas, Barbara and Mazon, D and Delaunay, F and Usai, P and Delmond, JF},
  journal={IEEE Transactions on Plasma Science},
  volume={38},
  number={12},
  pages={3409--3418},
  year={2010},
  publisher={IEEE}
}

@article{Craciunescu2012FST,
  title={Phase congruency image classification for {MARFE} detection on {JET} with a carbon wall},
  author={Craciunescu, T and Murari, A and Tiseanu, I and Vega, J and JET-EFDA Contributors},
  journal={Fusion Science and Technology},
  volume={62},
  number={2},
  pages={339--346},
  year={2012},
  publisher={Taylor \& Francis}
}

@article{Albuquerque2011TPS,
  title   = {High-Speed Image Processing Algorithms for Real-Time Detection of {MARFEs} on {JET}},
  author  = {Portes de Albuquerque, M. and de Albuquerque, M. P. and Chacon, G. and de Faria, E. L. and Murari, A. and JET EFDA contributors},
  journal = {IEEE Transactions on Plasma Science},
  year    = {2012},
  volume  = {40},
  number  = {12},
  pages   = {3485--3492}
}

@article{Raissi2019JCP,
  title   = {Physics-informed neural networks: A deep learning framework for solving forward and inverse problems involving nonlinear partial differential equations},
  author  = {Raissi, M. and Perdikaris, P. and Karniadakis, G. E.},
  journal = {Journal of Computational Physics},
  year    = {2019},
  volume  = {378},
  pages   = {686--707},
  doi     = {10.1016/j.jcp.2018.10.045}
}

@article{Willard2022CSUR,
  title={Integrating Scientific Knowledge with Machine Learning for Engineering and Environmental Systems},
  author={Willard, Jared and Jia, Xiaowei and Xu, Shaoming and Steinbach, Michael and Kumar, Vipin},
  journal={ACM Computing Surveys},
  volume={55},
  number={4},
  pages={1--37},
  year={2023},
  doi={10.1145/3514228},
  publisher={Association for Computing Machinery}
}

@inproceedings{Chen2018NeurIPS,
  title={Neural ordinary differential equations},
  author={Chen, Ricky TQ and Rubanova, Yulia and Bettencourt, Jesse and Duvenaud, David K},
  booktitle = {Advances in Neural Information Processing Systems},  
  volume={31},
  year={2018},
  address = {Montréal, Canada}
}

@inproceedings{serban2005cvodes,
  title     = {CVODES, the Sensitivity-Enabled ODE Solver in SUNDIALS},
  author    = {Serban, Radu and Hindmarsh, Alan C.},
  booktitle = {Proceedings of IDETC/CIE 2005, ASME International Design Engineering Technical Conferences},
  address   = {Long Beach, CA, USA},
  year      = {2005},
  note      = {Adjoint module uses checkpointing and Hermite interpolation},
}

@inproceedings{Rubanova2019NeurIPS,
  title={Latent ordinary differential equations for irregularly-sampled time series},
  author={Rubanova, Yulia and Chen, Ricky TQ and Duvenaud, David K},
  booktitle = {Advances in Neural Information Processing Systems},
  volume={32},
  year={2019},
  address = {Vancouver, Canada}
}

@inproceedings{Kidger2020NeurIPS,
  title={Neural controlled differential equations for irregular time series},
  author={Kidger, Patrick and Morrill, James and Foster, James and Lyons, Terry},
  booktitle = {Advances in Neural Information Processing Systems},
  volume = {33},
  year = {2020},
  address = {Virtual}
}

@article{GonzalezGanzabal2024FED,
  title   = {Advancing {MARFE} detection in {JET's} operational camera videos through Machine Learning techniques},
  author  = {Gonz{\'a}lez Ganz{\'a}bal, A. and Ratt{\'a}, G. A. and Gadariya, D. and Dormido-Canto, S.},
  journal = {Fusion Engineering and Design},
  year    = {2024},
  volume  = {205},
  pages   = {114534},
  doi     = {10.1016/j.fusengdes.2024.114534}
}

@article{Hu2023CPB,
  title={Prediction of multifaceted asymmetric radiation from the edge movement in density-limit disruptive plasmas on experimental advanced superconducting tokamak using random forest},
  author={Hu, Wenhui and Hou, Jilei and Luo, Zhengping and Huang, Yao and Chen, Dalong and Xiao, Bingjia and Yuan, Qiping and Duan, Yanmin and Hu, Jiansheng and Zuo, Guizhong and others},
  journal={Chinese Physics B},
  volume={32},
  number={7},
  pages={075211},
  year={2023},
  publisher={IOP Publishing}
}

@article{KatesHarbeck2019Nature,
  title={Predicting disruptive instabilities in controlled fusion plasmas through deep learning},
  author={Kates-Harbeck, Julian and Svyatkovskiy, Alexey and Tang, William},
  journal={Nature},
  volume={568},
  number={7753},
  pages={526--531},
  year={2019},
  publisher={Nature Publishing Group UK London}
}

@article{Peluso2019RSI,
  title={A comprehensive study of the uncertainties in bolometric tomography on {JET} using the maximum likelihood method},
  author={Peluso, E and Craciunescu, T and Murari, A and Carvalho, P and Gelfusa, M and JET, Contributors},
  journal={Review of Scientific Instruments},
  volume={90},
  number={12},
  year={2019},
  publisher={AIP Publishing}
}

@article{youden1950index,
  title={Index for rating diagnostic tests},
  author={Youden, William J},
  journal={Cancer},
  volume={3},
  number={1},
  pages={32--35},
  year={1950},
  publisher={Wiley Online Library}
}

@article{Bernert2014RSI,
  title={Application of AXUV diode detectors at {ASDEX} Upgrade},
  author={Bernert, M and Eich, T and Burckhart, A and Fuchs, JC and Giannone, L and Kallenbach, A and McDermott, RM and Sieglin, B and ASDEX Upgrade Team and others},
  journal={Review of scientific Instruments},
  volume={85},
  number={3},
  year={2014},
  publisher={AIP Publishing}
}

@article{Craciunescu2023PhysScr,
  title={Maximum likelihood bolometry for {ASDEX} upgrade experiments},
  author={Craciunescu, Teddy and Peluso, Emmanuele and Murari, Andrea and Bernert, Matthias and Gelfusa, Michela and Rossi, Riccardo and Spolladore, Luca and Wyss, Ivan and David, Pierre and Henderson, Stuart and others},
  journal={Physica Scripta},
  volume={98},
  number={12},
  pages={125603},
  year={2023},
  publisher={IOP Publishing}
}

@article{Losada2020NME,
  title={Observations with fast visible cameras in high power Deuterium plasma experiments in the {JET} {ITER-like} wall tokamak},
  author={Losada, Ulises and Manzanares, A and Balboa, I and Silburn, S and Karhunen, J and Carvalho, Pedro J and Huber, A and Huber, V and Solano, Emilia R and De La Cal, E and others},
  journal={Nuclear Materials and Energy},
  volume={25},
  pages={100837},
  year={2020},
  publisher={Elsevier}
}

@article{Lotte2010RSI,
  title={Wall reflection issues for optical diagnostics in fusion devices},
  author={Lotte, Ph and Aumeunier, MH and Devynck, P and Fenzi, C and Martin, V and Trav{\`e}re, JM},
  journal={Review of Scientific Instruments},
  volume={81},
  number={10},
  year={2010},
  publisher={AIP Publishing}
}

@article{Carr2019RSI,
  author  = {M. Carr and A. Meakins and S. A. Silburn and J. Karhunen and M. Bernert and et al.},
  title   = {Physically principled reflection models applied to filtered camera imaging inversions in metal walled fusion machines},
  journal = {Review of Scientific Instruments},
  year    = {2019},
  volume  = {90},
  number  = {4},
  pages   = {043504},
  doi     = {10.1063/1.5092781}
}

@article{Marini2023RSI,
  title={The fast camera (Fastcam) imaging diagnostic systems on the {DIII-D} tokamak},
  author={Marini, Claudio and Boedo, JA and Hollmann, EM and Chousal, L and Mills, J and Popovi{\'c}, Z and Bykov, I},
  journal={Review of Scientific Instruments},
  volume={94},
  number={5},
  year={2023},
  publisher={AIP Publishing}
}

@article{Hollmann2015PoP,
  title={Status of research toward the {ITER} disruption mitigation system},
  author={Hollmann, EM and Aleynikov, PB and F{\"u}l{\"o}p, T{\"u}nde and Humphreys, DA and Izzo, VA and Lehnen, M and Lukash, VE and Papp, Gergely and Pautasso, G and Saint-Laurent, F and others},
  journal={Physics of Plasmas},
  volume={22},
  number={2},
  year={2015},
  publisher={AIP Publishing}
}

@article{adams2007bocd,
  title   = {Bayesian Online Changepoint Detection},
  author  = {Adams, Ryan P. and MacKay, David J. C.},
  journal = {arXiv preprint arXiv:0710.3742},
  year    = {2007}
}

@article{stroth2022model,
  title={Model for access and stability of the X-point radiator and the threshold for {MARFES} in tokamak plasmas},
  author={Stroth, U and Bernert, M and Brida, D and Cavedon, M and Dux, R and Huett, E and Lunt, T and Pan, O and Wischmeier, M and ASDEX Upgrade Team and others},
  journal={Nuclear Fusion},
  volume={62},
  number={7},
  pages={076008},
  year={2022},
  publisher={IOP Publishing}
}

@article{duan2022progress,
  title={Progress of HL-2A experiments and HL-2M program},
  author={Duan, XR and Xu, M and Zhong, WL and Liu, Y and Song, XM and Liu, DQ and Wang, YQ and Lu, B and Shi, ZB and Zheng, GY and others},
  journal={Nuclear Fusion},
  volume={62},
  number={4},
  pages={042020},
  year={2022},
  publisher={IOP Publishing}
}

@article{dempster1977maximum,
  title={Maximum likelihood from incomplete data via the {EM} algorithm},
  author={Dempster, Arthur P and Laird, Nan M and Rubin, Donald B},
  journal={Journal of the royal statistical society: series B (methodological)},
  volume={39},
  number={1},
  pages={1--22},
  year={1977},
  publisher={Wiley Online Library}
}

@book{mclachlan2008algorithm,
  title={The EM algorithm and extensions},
  author={McLachlan, Geoffrey J and Krishnan, Thriyambakam},
  year={2008},
  publisher={John Wiley \& Sons}
}

@article{blanchard2021accurately,
  title={Accurately computing the log-sum-exp and softmax functions},
  author={Blanchard, Pierre and Higham, Desmond J and Higham, Nicholas J},
  journal={IMA Journal of Numerical Analysis},
  volume={41},
  number={4},
  pages={2311--2330},
  year={2021},
  publisher={Oxford University Press}
}

@inproceedings{kendall2018multi,
  title={Multi-task learning using uncertainty to weigh losses for scene geometry and semantics},
  author={Kendall, Alex and Gal, Yarin and Cipolla, Roberto},
  booktitle={Proceedings of the IEEE Conference on Computer Vision and Pattern Recognition},
  pages={7482--7491},
  year={2018},
  address = {Salt Lake City, UT, USA}
}

@article{Spolladore2023FED,
  title   = {Detection of {MARFEs} using visible cameras for disruption prevention},
  author  = {Spolladore, L. and Rossi, R. and Wyss, I. and Gaudio, P. and Murari, A. and Gelfusa, M. and Abate, D. and Abid, N. and Abreu, P. and Salzedas, F. and {JET Contributors}},
  journal = {Fusion Engineering and Design},
  volume  = {190},
  pages   = {113507},
  year    = {2023},
  doi     = {10.1016/j.fusengdes.2023.113507}
}

\end{document}